\documentclass[a4paper,fleqn,]{cas-sc}

\usepackage{caption}
\usepackage{float}
\usepackage[numbers]{natbib}
\usepackage{multicol} % <-- ADD THIS LINE
\def\tsc#1{\csdef{#1}{\textsc{\lowercase{#1}}\xspace}}
\tsc{WGM}
\tsc{QE}

\begin{document}
\let\WriteBookmarks\relax
\def\floatpagepagefraction{1}
\def\textpagefraction{.001}

\shorttitle{}    

%\shortauthors{Damianova \& Berrezueta-Guzman}  

\title [mode = title]{The Anatomy of a Successful Student Scrum Team: Motivation, Personalities, and Academic Adaptation}  




\author[1]{Nadia Damianova}%[<options>]

%\cormark[1]


%\ead{nadia.damianova@tum.de}


\credit{Methodology, Writing, Surveys, Data Collection}

\affiliation[1]{organization={Technical University of Munich},
            addressline={}, 
            city={Heilbronn},
            country={Germany}}

\author[1]{Santiago Berrezueta-Guzman}[orcid=0000-0001-5559-2056]

% Footnote of the second author

% Email id of the second author
\ead{s.berrezueta@tum.de}

%\ead[url]{000000}

\credit{Conceptualization, Writing, Supervision}

%\affiliation[1]{organization={Technical University of Munich},
 %           addressline={}, 
 %           city={Heilbronn},
 %           country={Germany}}

\cortext[1]{Corresponding author}

%\fntext[1]{00000}


\begin{abstract}
Agile methods, and Scrum in particular, are widely taught in software engineering education; however, there is limited empirical evidence on how these practices function in long-running, student-led projects under academic and hybrid work constraints. This paper presents a year-long case study of an eight-person student development team tasked with designing and implementing a virtual reality game that simulates a university campus and provides program-related educational content. We analyze how the team adapted Scrum practices (sprint structure, roles, backlog management) to fit semester rhythms, exams, travel, and part-time availability, and how communication and coordination were maintained in a hybrid on-site/remote environment. Using qualitative observations and artifacts from Discord, Notion, and GitHub, as well as contribution metrics and a custom communication effectiveness index (score: 0.76/1.00), we evaluate three dimensions: (1) the effectiveness of collaboration tools, (2) the impact of hybrid work on communication and productivity, and (3) the feasibility of aligning Scrum with academic timelines. Our findings show that (i) lightweight, tool-mediated coordination enabled stable progress even during remote periods; (ii) one-week sprints and flexible ceremonies helped reconcile Scrum with academic obligations; and (iii) shared motivation, role clarity, and compatible working styles were as critical as process mechanics. We propose practical recommendations for instructors and student teams adopting agile methods in hybrid, project-based learning settings.
\end{abstract}


%\item Year-long case study of an eight-person student Scrum team building a VR campus game.
%\item Mixed-method analysis of Discord, Notion, and GitHub artefacts in a hybrid setting.
%\item Communication effectiveness index (EI = 0.76) operationalizes team coordination quality.
%\item Concrete patterns for adapting Scrum sprints and ceremonies to academic calendars.
%\item Shows how motivation, role clarity, and introvert-leaning personalities shape success.



\maketitle
\begin{multicols}{2}

\section{Introduction}\label{I}

Scrum has become the most widely adopted agile methodology nowadays, with approximately 75\% of organizations using it or a hybrid version \cite{alami2022scrum}. At its core, Scrum emphasizes self-organizing, cross-functional teams that work in short iterations called sprints, which typically last one to two weeks or less. The framework consists of three primary roles: the product owner, the scrum master, and the development team. 
Research has demonstrated that Scrum improves software quality in two main ways: it encourages teamwork, trust, accountability, and openness, and it enhances processes through repeated development cycles and regular reviews with adjustments \cite{alami2022scrum, kadenic2023mastering}. Given its widespread use, it is essential to introduce Scrum to students before they enter the professional field. Today, Scrum extends beyond software engineering, finding applications in almost any field of work \cite{hidalgo2019adapting}.

In higher education, Scrum and other agile practices are increasingly integrated in project-based learning to structure teamwork and provide students with a realistic experience in modern development processes. Prior work reports benefits such as improved coordination, transparency of responsibilities, and increased motivation \cite{persson2011use, fernandes2021improving}. At the same time, educators consistently note tensions between standard Scrum practices and academic realities: courses are bounded by semesters, students juggle multiple modules and exam periods, and project work is often part-time and unevenly distributed over the year \cite{masood2018adapting}. These constraints are further complicated by the widespread adoption of hybrid and remote work models, which also shape how future graduates will collaborate in professional settings.

Despite growing interest in agile methods in education, three significant gaps remain. First, many reports focus on single-semester courses or short projects, providing limited longitudinal evidence on how student Scrum teams function over longer timeframes \cite{chowdhury2025agile, persson2011use}. Second, most studies either assume fully remote teams or treat remote work as temporary, which offers very little insight into hybrid collaboration as a normal mode of operation \cite{khanna2024hybrid, christensen2025hybrid}. Third, existing work rarely connects process mechanics, such as sprints, ceremonies, and backlogs, with human factors like motivation, personality, and working styles, despite these aspects being known to influence software team performance in industry settings \cite{beecham2008motivation, cruz2015forty}. As a result, we still lack empirical accounts of how student teams adapt Scrum in practice over a period of many months under academic and hybrid work constraints, and whether collaboration tools support this adaptation.

This study addresses these gaps through a year-long case study of a student development team tasked with creating a virtual reality (VR) game. The project involved designing and implementing an immersive simulation of a university campus that also delivers program-related educational content. An eight-person team adopted a Scrum-inspired development process, which utilized one-week sprints, a product backlog managed in Notion, and regularly held sprint reviews in a hybrid on-site/remote format. Coordination and day-to-day communication relied heavily on three collaboration tools: Discord (for synchronous and asynchronous communication), Notion (for planning, documentation, and task tracking), and GitHub (for version control and code review). The authors participated in and observed this project over the course of a full year, from December 2024 to November 2025.

We combine qualitative observations with artifacts and metrics from these tools, including Discord conversations, Notion task records and documentation, and GitHub contribution data, to analyze how agile practices were implemented in this educational setting. In particular, we define a communication effectiveness index that operationalizes coordination quality along four dimensions (activity, responsiveness, clarity, and outcome), yielding a score of 0.76/1.00 for the team. We then interpret this score together with commit patterns, workload distribution across milestones, and narrative accounts of team dynamics, motivation, and personalities.

Through this paper, we aim to answer the following questions:

\textit{\textbf{RQ1:} How effective are collaboration tools in supporting coordination in a long-running student Scrum team?}

\textit{\textbf{RQ2:} Does a hybrid working environment hinder communication and productivity in a student Scrum team?}

\textit{\textbf{RQ3:} How can Scrum be adapted to align with academic timelines and student constraints?}

Our contributions are threefold. First, we present a longitudinal case study of an eight-person student Scrum team working for a whole year on a VR campus game under hybrid on-site/remote conditions, providing rich, empirical detail on how process, tools, and human factors interacted over time. Second, we propose and evaluate a communication effectiveness index that integrates Discord activity, Notion task tracking, and GitHub commits into a simple, interpretable measure of coordination quality. Third, we identify concrete patterns for adapting Scrum to academic constraints, such as one-week sprints, planned "pause weeks" around exams, flexible ceremonies, and role mappings, and relate these patterns to students’ motivation and personality-influenced working styles.

The remainder of this paper is organized as follows. Section~\ref{RW} reviews related work on Scrum in higher education, agile adaptations beyond industry, hybrid and remote teamwork, human factors in software engineering, and collaboration tools, positioning our study within this broader landscape. Section~\ref{PS} details the research design, including the VR campus game context, team roles, development methodology, data sources, and analysis procedures. Section~\ref{sec:results} presents the empirical results for our three research questions, and Section~\ref{F} synthesizes these results into broader findings on adapting Scrum to academic constraints, collaboration in hybrid settings, and the role of motivation and personality in team dynamics. Section~\ref{TV} then examines threats to validity, while Section~\ref{D} discusses how our findings relate to existing work on agile methods and human factors. Section~\ref{IM} derives implications for instructors, student teams, and researchers. Finally, Section~\ref{C} concludes the paper by summarizing key insights and outlining directions for future work on agile, project-based learning in hybrid environments.

\section{Related Work}\label{RW}

This section aims to review past research that helps place this study in the broader context of software engineering collaboration, agile methods, and student learning environments. It examines how frameworks such as Scrum and Agile have been applied in education to enhance teamwork, motivation, and learning. It also considers studies on remote and hybrid work, human factors, and how personality affects team performance. Additionally, we talk about the tools GitHub, Notion, and Discord, as they were the primary platforms that supported collaboration and communication throughout the project. 

\subsection{Scrum in higher education}
Many studies investigate the application of Scrum or Scrum-like practices in undergraduate or graduate project courses. Persson et al. demonstrate that Scrum can help with project-based learning and increase visibility of tasks and responsibilities in student teams \cite{persson2011use}. Other empirical evaluations report positive effects on team coordination, process transparency, and learning outcomes. However, outcomes depend strongly on instructor monitoring and students’ prior experience with team-based development \cite{fernandes2021improving}. Case studies also highlight tensions between Scrum’s sprints and academic scheduling constraints (semester boundaries, assignment deadlines) \cite{masood2018adapting}, which suggests adaptation is often necessary. 

Many researchers have explored how Scrum, originally an agile practice from the software industry, can be effectively employed in project-based learning (PBL) settings in higher education. Fernandes et al., conducted a case study in an engineering program, where they compared two editions of PBL courses using Scrum with differing characteristics (number of teams, distribution of Scrum roles, etc.), where findings conclude that Scrum seems to help with keeping the project on track, improving task management, and team motivation \cite{carvalho2018making}. 

Research is also being conducted into Scrum’s effects on soft skills and how students perceive Scrum in the context of PBL. A recent study with third-year university students shows that using Scrum in PBL increases motivation, collaboration, communication, and problem-solving compared to more traditional methods \cite{vcisar2025impact}. The mechanisms cited include daily stand-ups, retrospectives, and iterative cycles, which help expose students to what works and what does not early on. Similarly, Aragonés-Jericó et al., reported that students value Scrum’s ability to promote interactivity, commitment, and adaptability \cite{aragones2022agile}. In the study, students found that learning by doing, receiving frequent feedback, and working under realistic pressures were useful.

\subsection{Agile adaptations in non-industrial contexts}

While Agile frameworks were initially designed for software development, their principles, such as iterative progress, collaboration, and flexibility, have proven beneficial in non-industrial contexts \cite{gustavsson2016benefits}. In sectors such as marketing, healthcare, and education, organizations have adapted Agile principles to enhance collaboration, responsiveness, and efficiency \cite{umezurike2025review, saleh2024agile, rajagopalan2025agile}. In marketing, for example, Agile practices have enabled quicker adaptation to market trends and improved customer engagement \cite{umezurike2025review}. Similarly, in healthcare, Agile methodologies have been employed to organize processes more effectively and improve patient care by strengthening the communication among teams \cite{saleh2024agile}. In academic environments, these methodologies can foster teamwork and adaptability \cite{rajagopalan2025agile}, which are recognized as essential skills for students engaged in projects. 

However, the direct application of Scrum in academic settings often requires significant adaptations. Traditional Scrum practices, such as daily stand-ups and fixed-length sprints, may not align with the academic calendar or the part-time commitment of students. Reynolds et al., suggest that educators can modify Scrum's structure to better fit the educational context, such as adjusting sprint lengths to match semester timelines and incorporating asynchronous communication tools to accommodate students' varying schedules \cite{reynolds2023scrum}. The biggest strength of Agile methodologies is their inherent flexibility, which allows teams to adapt practices to suit their specific project needs \cite{vandersluis2014apply}. 

\subsection{Hybrid and remote teamwork}
The move to hybrid and remote work has changed how software engineering operates, and the COVID-19 pandemic significantly accelerated this shift that researchers had already been observing for years \cite{nguyen2024work}. Recent studies reveal that hybrid working has become the dominant model in engineering \cite{bloom2022hybrid}. A 2024 systematic literature review found that 83 software companies reported more positive responses than negative regarding hybrid work, primarily due to better work-life balance and geographical flexibility \cite{khanna2024software}. Khanna et al. investigated hybrid and remote work in agile teams and found a major adoption of remote work when such options were provided. \cite{khanna2024hybrid}. However, companies commonly implement guidelines regarding the amount of remote work allowed \cite{ask2024hybrid}.

Recent work shows that hybrid and remote work change how teams communicate, coordinate, and use tools \cite{de2024meetings, khanna2024hybrid}. Research of student and industry teams finds that hybrid setups increase coordination challenges and reduce casual “\textit{hallway}” interactions \cite{ask2024hybrid, handke2024hybrid}. To adapt, teams rely more on scheduled meetings, detailed documentation, and collaboration platforms. Findings on hybrid Agile teams are mixed - these arrangements improve flexibility but also create fragmented communication and hidden dependencies \cite{priyanka2025analyzing}.

\subsection{Human factors in software engineering}

Human factors should not be considered irrelevant in software engineering, as they significantly impact productivity and ultimately, the quality of the product that is built. Several literature reviews argue that cognition, motivation, personality, and social processes are central to software outcomes and must be treated as high-priority concerns in such practices \cite{beecham2008motivation, lenbergabehavioral, zolduoarrati2023secondary}. As noted by other experienced engineers, while technologies and tools continuously evolve, the human element remains constant and essential to the success of any project \cite{millerhuman}. 

Today, we have advanced tools like artificial intelligence and other domain-specific specializations. Still, if we look beyond the technology and compare it to software development 25 years ago, the core aspect remains the same: a team collaborating, sharing their work, and progressing together. Therefore, the effectiveness of software development ultimately depends on the people involved, their skills, collaboration, and understanding, rather than solely on the ever-changing technical environment \cite{millerhuman}. 

A large body of work shows that human and socio-psychological factors (motivation, role clarity, communication quality, personality traits) substantially influence software team outcomes. Empirical coordination theories show that how teams communicate affects project structure and performance, a link confirmed by both academic and industry research \cite{herbsleb2003empirical}. Systematic reviews highlight that clear roles, psychological safety, and good alignment between communication tools and software structures are key for effective teamwork \cite{santana2025psychological}. Recent research also encourages the use of insights from psychology and organizational behavior to better understand what motivates developers, how they manage cognitive load, and how collaboration tools support their work \cite{graziotin2018happens, culas2025advancing}.

While technologies and tools continuously evolve, the human element remains constant and essential to the success of any project. This human element encompasses both hard skills (technical expertise) and soft skills, such as communication quality, teamwork, adaptability, and conflict resolution. These soft skills are particularly critical in agile environments, such as Scrum, where frequent and honest feedback, as well as self-organization, are essential for success. Studies highlight that clear roles, psychological safety, and good alignment between communication tools and software structures are key for effective teamwork \cite{andree2025} - all of which rely on a team's collective soft skills.

\subsection{Collaboration Tools}

Software engineering teams have become overly dependent on collaboration tools, with these platforms now forming the essential infrastructure of modern development. The research reveals that collaboration tools are not merely conveniences, but fundamental necessities that enable teams to function effectively in today's agile work environment \cite{hussein2025collaboration}. The dependency on collaboration tools stems from the inherently collaborative nature of software engineering itself. Research shows that developers spend the vast majority of their working hours on collaborative activities rather than solo coding \cite{sarma2005survey}. A 2011 study by Goncalves et al. found that developers spend only about 9\% of their time writing code, with the rest allocated to collaboration (45\%) and seeking information (32\%) \cite{meyer2019today}. Likewise, broader empirical estimates indicate that nearly 70\% of a software engineer’s time is spent on collaborative activities \cite{sarma2005survey}.
In this study, however, we will mostly pay attention to the following collaboration tools:

\subsubsection{Notion:}

Notion’s synced database and embedding capabilities (such as GitHub integration and the linking of documents or project boards) enable software engineering teams to track every kind of metadata. This functionality enhances overall project visibility without requiring all contributors to use the same issue tracker. Prior work on collaboration in software engineering emphasizes that tighter integration between artifacts (docs, boards, code links) and easy-to-use knowledge bases improves coordination and reduces overhead in teams - conditions that Notion’s all-in-one workspace is explicitly designed to support \cite{whitehead2007collaboration, goldbergstory}. 

Recent research has also begun to treat Notion as an example of a visual and flexible workspace, where users can store and coordinate shared project artifacts \cite{redmond2023project}. However, currently, there is a lack of research that measures the effects of Notion's use on software engineering outcomes, such as development outcomes, code quality, maintenance overhead, and error rates.

\subsubsection{Discord:}

Studies that analyze Discord conversations reveal that developer and user discussions hosted there contain program-relevant information that can aid in program comprehension and knowledge sharing \cite{raglianti2022using}. Open datasets, such as DISCO, enable the analysis of real project-community dialog \cite{subash2022disco}. Classroom and educational case studies report that Discord’s social features (voice, persistent channels) support peer learning, student collaboration, and community building \cite{alghamdi2025bridging}. 

At the same time, research cautions that chat platforms introduce frequent interruptions, searchability problems, and security/abuse concerns - factors that may harm concentrated development work or increase maintenance overhead if not governed \cite{pozdniakov2025misunderstandings}. Taken together, the literature suggests Discord functions as a flexible “third-place” for project interaction \cite{kim2025discord}. However, the causal evidence linking Discord use to concrete software engineering outcomes remains sparse: researchers have demonstrated the existence of content and its support for comprehension/learning, but there are few large-scale, controlled measurements of downstream engineering outcomes to date \cite{berrezueta2024interactive}.

\subsubsection{GitHub:}
GitHub not only helps developers manage code, but it also adds social and collaborative features, such as issues, pull requests, and discussions, that make it easier for teams to review code, report bugs, and coordinate development. Due to these features, GitHub has become one of the primary tools for software development, currently supporting millions of users and projects worldwide \cite{github2025}. 
Beyond its social features, GitHub enables team members to learn from each other’s code. Observing how others approach problems, structure their solutions, and prioritize different aspects of development provides access to learning more effective coding practices and sharing expertise \cite{dabbish2012social}.

\subsection{Personalities in software engineering}
Research spanning over 50 years has established that personality traits significantly influence software engineering outcomes, with studies showing measurable impacts on individual performance, team dynamics, and project success \cite{soomro2016effect}. According to Russo et al., approximately 33\% of the developers' cognitive engagement can be explained by personality traits, with openness to experience, conscientiousness, and low emotionality being particularly important predictors \cite{russo2022anecdote}. Multiple empirical studies have linked specific personality traits to concrete outcomes. 

Extraversion is positively correlated with team collaboration and software quality. Conscientiousness predicts project success and effective team performance. At the same time, high neuroticism has been associated with developer burnout and reduced creativity - different phases of the development lifecycle benefit from other personality types. Extroverts excel in requirements gathering and user-facing roles, while introverts perform better in independent coding tasks. Software engineers as a population score higher on need for cognition than the general population, suggesting that enjoyment of cognitive effort is a defining characteristic of successful developers \cite{zahl2023teamwork, capretz2015influence, endriulaitiene2021team, amin2020impact}.

The three primary personality models used in research are MBTI, Big Five, and HEXACO \cite{barroso2017influence}. They have each revealed that software engineers differ systematically from the general population, scoring higher on traits like thinking, judging, and need for cognition \cite{capretz2003personality}. Research interest has surged dramatically, with over 70\% of personality studies in software engineering published after 2002, indicating growing recognition of human factors in the field \cite{cruz2015forty}. 

Practical applications include using personality assessments for recruitment, forming balanced teams with diverse personalities, and matching individuals to roles that align with their traits. Research consistently shows that organizations ignoring personality factors miss opportunities to optimize team composition, reduce burnout, and improve project outcomes \cite{hidellaarachchi2024impact}.
\subsection{Gaps in the research}
Despite broad coverage, several gaps remain:

\begin{itemize}
    \item Longitudinal evidence in educational contexts: Many Scrum in education reports do not document long-term projects, but rather study semestral/course-length student software engineering workflows.

    \item Research beyond major platforms: While existing literature increasingly examines tools such as Slack, Git, and Jira, there is not so much empirical work assessing newer hybrid collaboration stacks. In particular, research on Notion as an integrated space for project management and planning, as well as on Discord as a platform for both synchronous and asynchronous team coordination, remains limited, especially within student team contexts. Recent surveys of collaboration tools also highlight the rapid pace of platform evolution and the lack of independent empirical evaluations \cite{jackson2022collaboration}. 

    \item Evidence on hybrid sprint rhythms and assessment alignment: There is little consensus on how to adapt sprint length, deliverables, and assessment strategies so Scrum practices align with semester schedules and learning assessment goals; the existing work provides recommendations but lacks controlled comparisons \cite{santana2017scrum}. 
\end{itemize}

The literature suggests that Scrum and Agile practices can help structure student projects and enhance coordination. However, their effectiveness relies on a more thoughtful adaptation to specific contexts. Factors such as institutional constraints (e.g., semester timelines), tool selection (including version control systems and management platforms), and socio-technical alignment (ensuring that communication channels reflect dependency structures) all shape outcomes. These insights underscore the importance of long-term studies that examine the relationship between tools, individual traits (such as motivation and personality), communication styles, and project organization within these teams.

\section{Research Design}\label{PS}
In this section, we describe the project, the student team and their roles, and the adapted Scrum development process used to structure the work.

\subsection{Research Questions}
We focus on the three research questions presented in Section~\ref{RW}:

\textbf{RQ1: How effective are collaboration tools in supporting coordination in a long-running student Scrum team?}

As we discussed previously, existing work has primarily examined widely adopted tools (Git, Jira, or Slack) in the context of course-based projects. Research evidence on newer hybrid collaboration platforms, particularly those investigated in this paper, including Notion, Discord, and GitHub, remains limited. \textbf{RQ1} aims to find the extent to which these tools can enhance software development by examining our student team case. We address this question by analyzing usage logs from all three platforms and by constructing a communication effectiveness index that captures activity, responsiveness, clarity, and task completion.

\textbf{RQ2: Does a hybrid working environment hinder communication and productivity in a student Scrum team?}

\textbf{RQ2} investigates how alternating between on-site and remote work affects team communication patterns and development output. To answer this question, we combine qualitative observations with quantitative indicators such as commit frequency, task activity across milestones, and the temporal distribution of Discord interactions. Prior research reports both benefits (flexibility, autonomy) and drawbacks (reduced informal contact, coordination overhead) of hybrid and remote work in software development teams. Still, there is little longitudinal evidence in educational contexts. 

\textbf{RQ3: How can Scrum be adapted to align with academic timelines and student constraints?}

Many studies recommend using Scrum in project-based learning, yet offer only high-level guidance on how to reconcile sprints, ceremonies, and roles with semester schedules, examination periods, and part-time student availability. \textbf{RQ3} targets this gap by analyzing how our team tailored Scrum practices to accommodate academic workloads, travel, and time-zone differences. The goal is not only to document the concrete adaptations observed in the project, but also to create suggestions for structuring similar student team methodologies in such settings.

\subsection{Project Context}

The project involves creating a navigable 3D virtual reality game of the Technical University of Munich (TUM) Campus in Heilbronn, designed for users to explore the campus environment in an immersive manner. Through this VR game, players can learn about the study program "Information Engineering" offered at TUM Campus Heilbronn. The development process spanned a year, from December 2024 to December 2025. A small student team was responsible for carrying the project through all stages of production, including design, development, testing, and evaluation. The team adopted a game development methodology inspired by standard software engineering practices and implemented the project using Unreal Engine 5. The project pursued two main objectives: an informational goal - to deliver an engaging and accurate virtual representation of the campus and study program; and a pedagogical goal - to examine how immersive, interactive environments can enhance learning through educational content.

\subsection{Team Roles}

The project's development team consisted of eight members in total: seven developers, including the project manager, and a project leader. Initially, the team consisted of only four developers, but as the project progressed and the workload increased, three additional developers were recruited to support the team's progress. In parallel with the game’s development, all team members also conducted research on topics related to the scope of the game (Unreal Engine 5, serious games, etc.). The team members were students aged between 20 and 23 years, all enrolled in the same academic program, which provided them with a shared set of technical and general knowledge, relevant to the story and scope of the project. Most members were familiar with each other before the project, which made the collaboration easier and the communication less tense.

\subsection{Development Methodology}
\begin{figure*}[h]
    \centering
    \includegraphics[width=\textwidth]{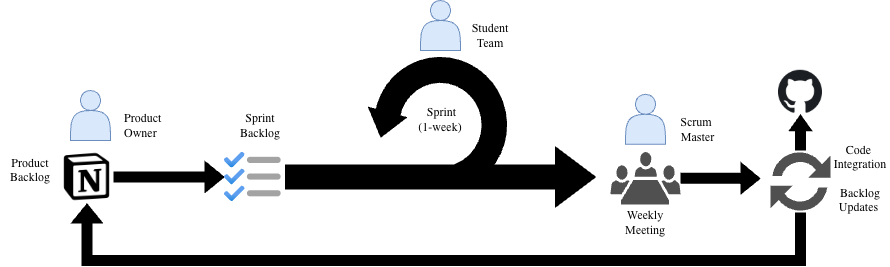}
    \caption{Scrum-based workflow used in the project.}
    \label{scrum}
\end{figure*}
The development process used for this project was based on a Scrum-like methodology, which emphasized iterative progress, flexibility for the students, and continuous feedback. Work was organized into one-week development cycles, or sprints, during which features were implemented and prepared for review. At the end of each sprint, deliverables were presented and evaluated in the team’s regular meetings to ensure consistent progression toward the project’s overall objectives.

Team meetings were conducted every week, either on-site or online. Given that all team members were international students, it was common for participants to travel home during semester breaks or holidays. As a result, the team needed to adapt its workflow to accommodate members working from different time zones and locations. Initially, meetings were held on Zoom, which provided a more formal environment for discussion. However, the team transitioned to Discord, as the platform allowed for more flexible and faster communication, as well as a more informal atmosphere. During meetings, newly developed features were discussed and tested. Constructive feedback was encouraged, and once a feature met the required standards and received team approval, it was integrated into the main project. 

Unlike traditional Scrum, where sprints typically last two to four weeks and emphasize client-driven deliverables, this project operated with one-week sprints to maintain a steady pace alongside academic workloads. The Product Backlog functioned as an ever-changing list of tasks and features maintained through Notion, similar to Scrum’s product backlog, which focuses on user stories and requirements. Each week, a subset of these tasks formed the Sprint Backlog, which guided the team’s immediate development goals. In alignment with Scrum roles, the Project Leader acted as the Product Owner, responsible for creating backlogs, setting priorities, and ensuring that development aligned with the project's overall goals. The Project Manager assumed the role of Scrum Master, taking the lead in communication, organizing sprint reviews, and helping the team overcome any development challenges that arose. The Development Team, composed of seven student developers, functioned as a self-organizing group that managed design, coding, and testing. While the team adopted Scrum’s iterative rhythm and feedback cycles, it replaced traditional ceremonies such as daily stand-ups and sprint planning with more flexible weekly reviews.

Figure \ref{scrum} visually summarizes how the development workflow connects all activities throughout each sprint. We can see how tasks moved from the product backlog in Notion into the weekly set of tasks (the sprint backlog), which the student development team implemented in one-week sprints. After each sprint, weekly meetings were led by the Scrum Master, during which features were reviewed and then integrated into the shared codebase (GitHub). Feedback and completed tasks were updated, and new tasks were added to the product backlog for the next sprint.
\subsection{Data Sources}

Data was collected manually through Discord and Notion logs, as well as through the built-in GitHub insights, including Commits, Code Frequency, Contributors, and other metrics. Additionally, team members were interviewed about their experiences, and although these insights were not formally documented, they also influenced the study’s findings. The authors, being active members of the team, also had the opportunity to closely observe the team’s dynamics. 

\subsubsection{Notion}
Throughout the project's duration, every task, meeting, and key decision was documented using Notion, which served as the central project management system. Tables were used to record all types of information, from creative components such as logo development and storyline progress to administrative tasks like contract management, budget tracking, and deadlines. This systematic documentation process significantly enhanced the project’s overall transparency, ensuring that all relevant information was easily accessible to every team member. It was also particularly beneficial for new developers joining the team, as they could quickly familiarize themselves with ongoing work and project history.

Notion functioned not only as a repository of information but also as a dynamic workspace for idea generation and team communication. Team members were encouraged to articulate their thoughts, brainstorm concepts, and draft proposals directly in the platform before sharing them for group discussion. Before each weekly meeting, all participants were required to update the meeting page with their weekly status, outline any challenges they had encountered, and specify their objectives or tasks for the upcoming week. This process facilitated clear communication by allowing every team member to reflect on their progress in writing before speaking it aloud. Additionally, the meeting documentation in Notion was very valuable for maintaining communication even when team members were missing. A designated minute-taker recorded essential points, which were then uploaded to the corresponding meeting page. This ensured that absent team members could easily review the meeting outcomes and remain informed about the project’s development.

\subsubsection{GitHub}

The project was stored in a GitHub repository, which further helped with data collection. Additionally, the use of GitHub Insights provided analytics regarding commit frequency, code ownership, and contribution patterns. These data metrics provided a quantitative perspective on team activity and collaboration, and later enabled the assessment of how the hybrid working setup influenced productivity.

\subsubsection{Discord}
Discord served as the primary platform for real-time communication among the developers. Dedicated channels were created to organize discussions by topic. The server was the primary space for daily collaboration on troubleshooting, scheduling meetings, and addressing random questions. Team members frequently shared game problems, screenshots, and quick updates, allowing issues to be resolved efficiently without the need for lengthy meetings. This immediate chat interaction proved highly effective in maintaining the workflow, especially during periods when spontaneous in-person discussions were not possible. Weekly development meetings were conducted through voice chat, often accompanied by screen sharing to review code, demonstrate features, or walk through technical challenges.

Since the collaboration of these tools is the primary focus of this paper, it's essential to understand their purpose in the study. Figure \ref{tools} provides an overview of the tools and data used, summarizing the key usages of the tools introduced in this section. 

\begin{center}
\begin{minipage}{\columnwidth}
    \centering
    \includegraphics[width=\columnwidth]{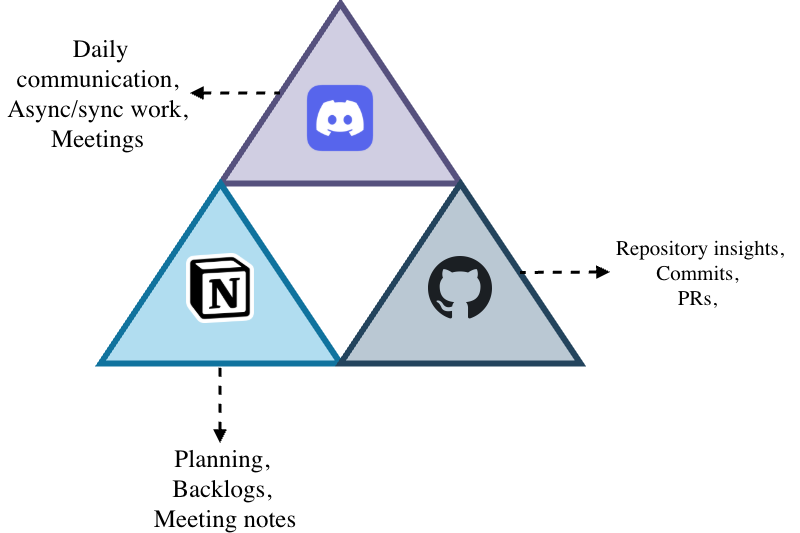}
    \captionof{figure}{Summary of collaboration tools used in the case study: Discord (top), Notion (left), GitHub (right).}
    \label{tools}
\end{minipage}
\end{center}

%%TODO
%I think it is missing the use of AI

\subsection{Data Analysis Procedures}

In this section, we describe how we processed raw data from Discord, Notion, and GitHub, and utilized qualitative information to support and validate our findings.

\subsubsection{Quantitative Analysis}

For Discord, we manually inspected all messages posted in the project channels during the study period and recorded them in a table. Table \ref{tab:discord} presents a summarized version of the table, provided for clarity. Messages were classified as project-related or not based on whether they referred to specific tasks, technical questions, planning, or decisions related to the game. Casual social exchanges and off-topic remarks were excluded. Questions were identified as messages explicitly requesting information, clarification, or help from other team members. For each question, we recorded whether it received an answer and the time elapsed until the first relevant reply. From these coded messages, we derived the four dimensions of the communication effectiveness index, which we discuss in greater detail in Section 5.2.2.

\begin{center}
\begin{tabular}{|l|p{2.5cm}|c|}
\hline
\textbf{Message Type} & \textbf{Description} & \textbf{Count} \\
\hline
Project-related & Tasks, technical questions, planning, polls & 233 \\
\hline
Off-topic & Social exchanges, unrelated remarks & 97 \\
\hline
Questions asked & Requests for info, clarification, or help & 54 \\
\hline
Questions answered & Questions that received a reply & 52 \\
\hline
\end{tabular}
\captionof{table}{Summary of Discord messages and their counts.}
\label{tab:discord}
\end{center}

For the analysis of commit activity, we used GitHub Insights to obtain monthly and weekly counts of commits. These values were exported and aligned with the project timeline to identify peaks and declines in development. Periods of predominantly remote work were marked based on the team’s meeting records, allowing us to visually compare activity across on-site and remote intervals.

\begin{figure*}[h]
    \centering
    \includegraphics[width=\textwidth]{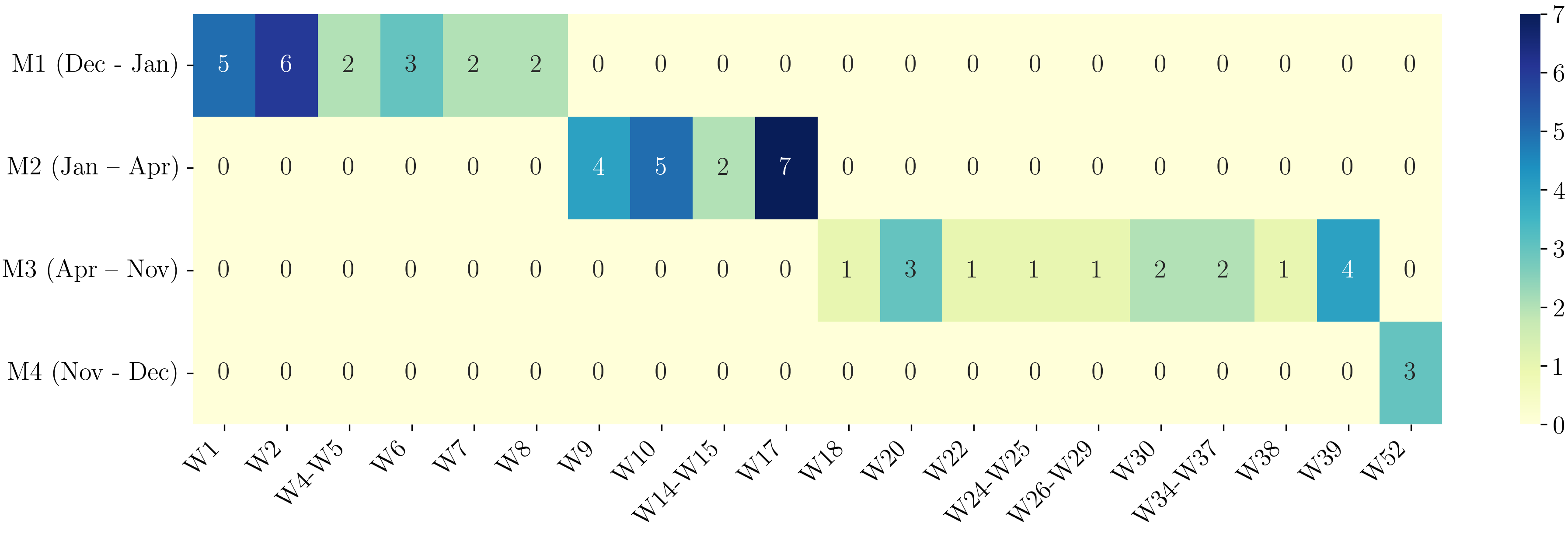}
    \caption{Task heatmap. Distribution of Tasks across a yearly timeline, broken down by weeks (x-axis) and monthly milestones (left y-axis).}
    \label{heatmap}
\end{figure*}

The task heatmap (Figure~\ref{heatmap}) was constructed from the Notion tasks database. Each task entry included an assignee, a creation date, an associated milestone, and a status. We grouped tasks by calendar week and milestone, and then counted the number of tasks assigned in each week–milestone combination. These counts were plotted as a heatmap, with rows representing milestones and columns representing weeks, to visualize the shift in workload over the year.

\subsubsection{Qualitative Analysis}

Qualitative data consisted of the authors’ observational notes from weekly meetings, the content of Discord conversations, and informal interviews with team members. These materials were not subjected to a complete formal thematic analysis; instead, we used them in an inductive and exploratory manner to identify recurring patterns related to motivation, working style, and perceptions of the hybrid setup.

Both authors reviewed meeting notes, selected Discord threads, and interview summaries to extract illustrative episodes (e.g., how the team handled exam periods, how decisions about tools and sprint structure were made). We then discussed these episodes to align our interpretations and to relate them to concepts from the literature, such as role clarity, working styles, and personal traits.

\section{Findings}\label{F}

The project provided a valuable learning experience for the students, as they were directly responsible for conceptualizing and developing the game. Rather than following predefined rules, the students took ownership of the creative process by making decisions regarding the storyline, gameplay mechanics, visual design, and essentially every other element of the game. Throughout the meetings, they collectively shaped the narrative and overall direction of the project, which enhanced both their technical and creative skills. Notably, there were no externally imposed rules or restrictions dictating what the game should be. This open-ended structure granted the students a degree of creative freedom, allowing them to experiment, share ideas, and reach consensus through discussion.

\subsection{Adapting Scrum to Academic Constraints}
Adapting the methodology to fit academic constraints created several challenges throughout the project. For example, it was occasionally necessary to postpone sprints by one or more weeks due to exam periods, holidays, or other academic obligations. In addition, as many team members were international students, periods of travel often resulted in limited access to necessary equipment, which further delayed the deliverables. Scheduling regular meetings also required careful coordination, as all participants needed to agree on specific times each week to accommodate their academic schedules.

The heatmap shown in Figure \ref{heatmap} provides a visual summary of how assigned tasks for the project were distributed over the course of a year, aligned with four key development milestones:

\begin{itemize}
    \item \textit{Milestone 1 – Clearly Defined Goals:} Tasks centered around establishing the world structure, brainstorming gameplay mechanics, drafting the story, and listing core systems.
    \item \textit{Milestone 2 – Prototype:} Focused on developing a playable test level, core controls, and a basic environment.
    \item \textit{Milestone 3 – Vertical Slice:} Implementing fully functional sublevels, integrating core systems, and adding placeholder art, audio, improved UI, and basic gameplay activities.
    \item \textit{Milestone 4 – Feature Freeze/Polish:} Dedicated to bug fixing, optimization, and overall user experience polish, i.e., the final refinement stage before completion.
\end{itemize}

The data visualized in the heatmap represents the number of tasks assigned per week within each milestone. While the numbers in each cell may seem modest, it is crucial to understand that many of these tasks were significant in terms of both complexity and effort required, which often demanded extended time commitments from students. The weekly task allocation was carefully tailored to accommodate the academic schedule and workload of the students. Notably, gaps in activity, such as those between weeks 10–14, 20–22, and 34–37, coincide with examination periods or retake sessions, which allowed students to prioritize their academic responsibilities during those times. The high task density during M1 indicates that pre-production required intense collaboration and decision-making before any development could proceed. Spikes such as the one in W17 (7 tasks) meant the students worked harder toward finishing the Prototype milestone. This shows deadline-driven surges rather than consistent workloads. Low task counts in e.g., M4 reflect that tasks were fewer in number but higher in complexity (e.g., bug fixes, optimizations).

\begin{figure*}[h]
    \centering
    \includegraphics[width=0.7\linewidth]{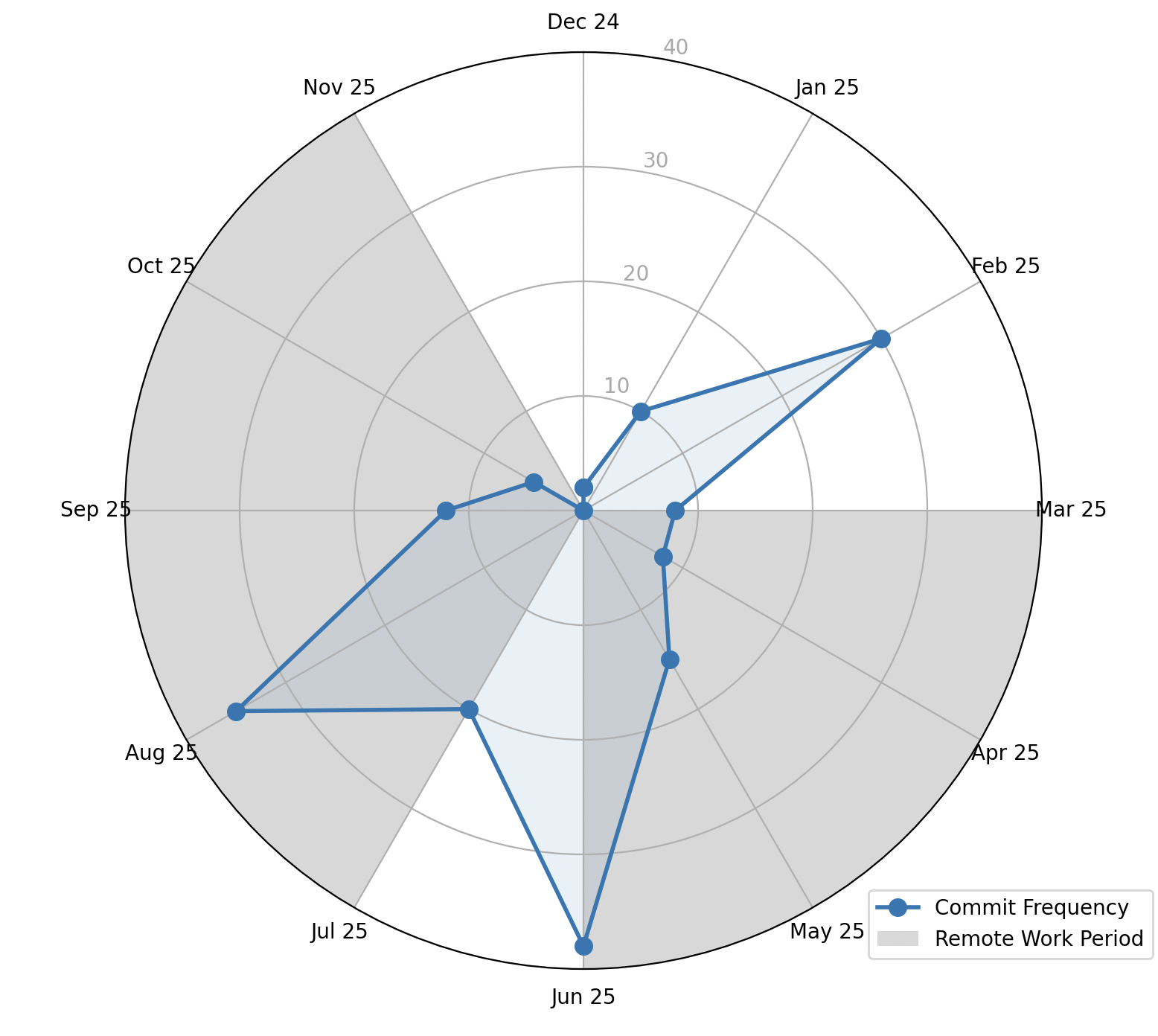}
    \caption{Commit Frequency graph. The blue line indicates all the commits made during a month; the grey areas show months of remote work.}
    \label{fig0}
\end{figure*}

\subsection{Communication and Coordination}

Communication within the group proved to be particularly effective. One of the key reasons for this was likely the shared background of the team members: most were of a similar age and enrolled in the same academic program, which was closely related to the game’s subject matter. This common foundation contributed to a strong sense of mutual understanding and collaboration, which in turn led to more efficient communication and a shared vision for the project’s development.

\subsubsection{Hybrid Work}\label{HW}

\begin{table*}[h]
\centering
\begin{tabular}{|p{3cm}|p{3.5cm}|p{7cm}|}
\hline
\textbf{Dimension} & \textbf{Collected Data} & \textbf{Thresholds} \\
\hline
Activity (A) & Message count and participation & 
$\geq 400$ \& >90\% members active = 1.0 \newline
250–399 \& >80\% members active = 0.8 \newline
150–249 \& moderate participation = 0.6 \\
\hline
Responsiveness (R) & Time between question and first reply, percentage of unanswered questions & 
< 10 min = 1.0 \& $\leq 5$\% unanswered questions \newline
11-20 = 0.8 \& $\leq 10$\% unanswered questions\newline
21-40 = 0.6 \& $\leq 15$\% unanswered questions \newline
41-60 = 0.4 \& $\leq 20$\% unanswered questions\newline
> 60 = 0.2 or \& >20\% unanswered questions\\
\hline
Clarity (C) & Sentiment and relevance of messages & 
$\geq 85\%$ = 1.0 \newline
75-84\% = 0.8 \newline
60-74\% = 0.6 \newline
<60\% = 0.4 \\
\hline
Outcome (O) & Tasks completed on time & 
$\geq 90\%$ = 1.0 \newline
75-89\% = 0.8 \newline
60-74\% = 0.6 \newline
40-59\% = 0.4 \newline
< 40\% = 0.2 \\
\hline
\end{tabular}
\caption{Pillars and Thresholds for Effectiveness Index Calculation}
\end{table*}

The commit frequency chart, illustrated in Figure \ref{fig0}, presents the count of monthly GitHub commits and identifies whether these were made in a remote or on-site setting. The data reveals fluctuations in developer activity, with notable peaks in February and June. Moderate levels of activity during April and May indicate consistent engagement, whereas declines in December, January, and October correspond to holiday schedules and transitional project phases, such as working on the storyline and design, instead of actively developing.

Notably, the comparison between remote and on-site work periods shows that remote work did not diminish productivity; in several cases, commit frequencies were sustained or even elevated during remote intervals. This suggests that hybrid work arrangements can still maintain stable development output.

From a communication perspective, these patterns imply that hybrid or remote conditions did not deplete the developers’ ability to coordinate effectively. The sustained commit rates indicate that teams were able to share information, manage version control, and utilize tasks efficiently through the digital communication tools. In hybrid environments, asynchronous communication channels may provide even greater clarity and reduce disruptions, enabling developers to contribute more autonomously.
Therefore, based on the observed data, the hybrid working environment appears not to hinder communication and may, in fact, enhance certain aspects of collaboration by encouraging structured, tool-mediated exchanges and enabling flexibility without sacrificing output.

\subsubsection{Efficacy of Discord}
We created an index to measure effectiveness, as illustrated in Table 1. The index was calculated to evaluate the effectiveness of the team communication in the \textit{Discord} server during the project period. The dataset included 384 project-related messages exchanged among seven active members, meaning full participation by all developers. The messages were manually checked and recorded in a table to extract the relevant information. For the Outcome (O) dimension, task deadlines were checked through the project's "Tasks" table in \textit{Notion} and then compared to \textit{GitHub} commits, to confirm they were (not) completed on time and consequently estimate the percentage of completed tasks on time. The average response time was 28.5 minutes, the proportion of unanswered questions was 4\%, and 70\% of tasks were completed on time. 

To interpret this data, each communication dimension was normalized on a 0–1 scale. With 384 messages and complete member participation, the activity score was assigned a value of 0.85 (since everyone remained active in the server, we added 0.05 to the percentage). The Responsiveness category (R) considered both response speed and the proportion of unanswered questions, resulting in an average score of 0.75 due to the moderate response delay. The Clarity component (C) was chosen as 0.80, given that participation was strong and few posts were ignored. The Outcome dimension (O) captured task timeliness, with a 70 percent on-time completion rate yielding an outcome score of 0.60. The effectiveness was calculated using the following weighted sum formula: 

$$\text{Effectiveness index (EI)}=w_1 A+w_2 R+w_3 C+w_4 O$$
Each category was weighted to reflect its importance according to the team: Activity ($w_1$ = 0.25), Responsiveness ($w_2$ = 0.30), Clarity ($w_3$ = 0.25), and Outcome ($w_4$ = 0.20). 
The result is a value between 0 and 1, where: 

\begin{itemize}
    \item \textit{score} > 0.75 = highly effective communication; 
    \item 0.50–0.75 = moderate efficacy;
    \item \textit{score} < 0.50 = potential communication inefficiency.
\end{itemize}

Substituting the weights into the formula:

\[
\text{EI} = 0.25 \cdot 0.85 + 0.30 \cdot 0.75 + 0.25 \cdot 0.80 + 0.20 \cdot 0.60 = 0.76
\]

The resulting score of 0.76 indicates a moderate to high level of communication effectiveness. This suggests a team dynamic characterized by strong engagement, active participation, and overall message clarity, though with mild delays in response times and some variability in task completion punctuality. To improve efficiency in future cycles, the team could prioritize faster response times during active collaboration windows, strengthen coordination to increase on-time task delivery beyond 85 percent, and introduce post-sprint sentiment or clarity checks to collect quality-related metrics. 

\subsection{Team Challenges}
In this section, we discuss the challenges that the team encountered during the development process.

\subsubsection{Motivation}
Motivation within the team was maintained through establishing clear milestones, which provided a sense of progress and purpose. Knowing that each sprint or task contributed to a specific goal, such as completing a playable demo or finalizing the game’s storyline, helped the students sustain engagement and focus throughout the project. Motivation was supported through the systematic use of Notion for task management. Each team member was assigned specific responsibilities with clearly defined steps to complete each week. This structured approach allowed everyone to understand precisely what was expected of them and provided a clear path toward achieving their objectives. By reducing ambiguity and ensuring that everyone had a manageable and organized workflow, the team avoided feelings of being lost or overwhelmed.

\subsubsection{Personalities in a team}
Within the team, individual personality traits influenced communication patterns, task ownership, and leadership dynamics. A few developers naturally took over leadership roles by guiding progress, clarifying requirements, and initiating discussions during sprint reviews. Leadership was often informal (rotating ownership of meetings or sprints) rather than strictly hierarchical, which helped the group stay flexible and inclusive. For all team members, the environment functioned as a balanced and organized collaboration, with tasks distributed according to skill and interest, and individuals expected to contribute autonomously to their respective areas of responsibility.

The team’s communication patterns suggested a predominantly introverted working style, with members preferring focused and independent work. They favored concise updates in meetings over lengthy discussions and limited side conversations that don’t directly move the project forward. This style had advantages - fewer distractions, deeper concentration, and higher-quality individual outputs. At the same time, the team demonstrated effective collaboration when problems arose. Problem-solving sessions were typically brief and focused on achieving a specific goal: a lead would frame the issue, relevant team members would present options, and the group would agree on an immediate next step. This aligns with our previous findings, as most software engineers tend to be introverted and prefer working alone, although they often have to face collaborative work.

\subsubsection{Challenges in a Hybrid Setting}
Time zones make hybrid teamwork more challenging, especially when team members have little overlap in their working hours. Different time zones disrupted communication, resulting in delays when some people were offline. Even a one-hour difference can make collaboration more challenging, complicate meeting schedules, and slow down overall project progress. To coordinate meetings across different time zones, the team always used polls to find times that worked for everyone. This ensured that all members, regardless of location, had the opportunity to participate. 

\section{Discussion and Future Work}\label{D}

In this section, we discuss how our findings address the proposed research questions, their validity, implications, and potential future work.

\subsection{Research Questions}\label{sec:results}
Table \ref{tab:tools-summary} provides an overview of the roles, observed benefits, and limitations of the three main collaboration tools we have discussed so far. The summary explains that the tools don’t operate in isolation, because Discord conversations depend on Notion documentation, GitHub activity depends on how tasks are defined, and so on. The limitations are some of our observations on why specific problems can arise in a team’s communication or task management, and how information can become fragmented, which can complicate the identification of communication patterns.

We will now discuss the results of the case study structured around the three research questions introduced in Section~\ref{PS} in greater detail. We compare our observations to existing work and make presumptions based on our findings in Section \ref{F}.

\textbf{RQ1 -- Effectiveness of Collaboration Tools}

Based on the Effectiveness Index and our overall observations, it can be concluded that these collaboration tools significantly improved the quality and organization of the students’ work. In their absence, the workflow would likely have been less structured, and new team members might have experienced confusion due to the lack of access to a centralized database of previous work, which is a feature supported by all three platforms. 

Prior research on collaboration platforms in higher education finds that such tools tend to increase behavioral engagement, for example, by encouraging more frequent contributions and making it easier for quieter students to participate asynchronously \cite{chandrasekar2011impact}. Within our team, the combination of text channels, issue tracking, and shared documents enabled members to contribute according to their own schedules and preferences, which helped maintain steady progress even when not everyone could attend meetings.

\begin{table*}[h]
\centering
\begin{tabular}{|p{1.5cm}|p{2.2cm}|p{5.2cm}|p{5.7cm}|}
\hline
\textbf{Tool} & \textbf{Primary Role} & \textbf{Observed Benefits} & \textbf{Observed Limitations} \\
\hline
Notion & Project management hub (backlog, milestones, meeting notes) &
Central, accessible overview of tasks, milestones, and responsibilities; easy onboarding of new members via documented history; written pre-meeting updates improved focus in discussions. &
Requires discipline to keep tasks and notes up to date; some overhead in manually linking tasks to code artefacts; limited built-in analytics compared to specialized issue trackers. \\
\hline
Discord & Day-to-day communication and hybrid meeting space &
Fast, informal coordination; low barrier for asking questions and sharing screenshots; dedicated channels kept topics organized; supported both synchronous (voice) and asynchronous (text) collaboration. &
Information can become fragmented across channels, the searchability of older discussions is limited, occasional delays occur when members are in different time zones, and not all informal decisions are systematically documented elsewhere. \\
\hline
GitHub & Version control and activity analytics &
Reliable record of code contributions; GitHub Insights enabled analysis of commit frequency and contributor patterns; facilitated peer learning by allowing members to inspect each other’s code. &
Commits do not capture all aspects of work (e.g., design, research); peaks and troughs in commit activity can be influenced by factors outside the repository (exams, holidays); linking commits to specific tasks requires consistent naming practices. \\
\hline
\end{tabular}
\caption{Summary of collaboration tools: roles, benefits, and limitations in the student Scrum team.}
\label{tab:tools-summary}
\end{table*}

\textbf{RQ2 -- Impact of Hybrid Work on Communication and Productivity}

In section \ref{HW}, we examined the metrics part of this question's findings and concluded that hybrid work does not have any advantages or disadvantages in terms of productivity. This could be due to the nature of the observed team, as they are students who have grown up with access to the internet. An interesting observation when inspecting messages from the Discord server is that in 80\% of the polls for choosing an online or on-site meeting, the team members chose the online option. This confirms the findings from recent research, which shows that Generation Z (Gen Z) prefers hybrid or flexible work and expects digitalized workplaces \cite{rani2025attracting}. 

This could be attributed to growing up online and having better digital skills than previous generations. Research on digital literacy indicates that individuals who are more confident with technology tend to perform better in digitalized work environments \cite{deschenes2024digital}. Because Gen Z tends to have higher digital comfort and expects collaboration via online tools, this can make it easier for them to adapt to remote collaboration. However, this does not automatically translate into higher productivity without effective management, clear goals, and manageable workloads.

From a communication perspective, the team appears to have benefited from a strong fit between their preferred communication style and the tools provided. A study of a software engineering course using Discord reports increased student interaction volume compared to office hours or forums \cite{bridson2022delivering}, suggesting that chat platforms can make seeking help and quick questions easier and less intimidating for students. In our case, the strong preference for online meetings and the persistent use of text channels mean that much of the team’s coordination and informal discussion is captured in writing rather than happening on-site. It appears that they preferred this working style, and from our observations, communication within the team remained effective even in a remote setting.

\textbf{RQ3 -- Adapting Scrum to Academic Timelines}
Our Scrum adaptation emphasizes iterative planning over detailed, fixed schedules. It lets teams continually reassess both sprint scope and overall objectives based on changing academic demands. Maintaining a flexible product backlog that evolves throughout the semester helps teams respond to surprises while keeping their long-term goals in sight.

Scrum can generally be adapted to student projects by making sprints and meetings more flexible to align with academic schedules and deadlines. They can be scheduled around exams, holidays, or busy weeks, with variable lengths or pause weeks during any deadlines. Traditional Scrum events can also be adjusted to better suit students’ needs. Daily stand‑ups, for example, can be replaced with brief online check‑ins, while the number of meetings may be reduced during hectic times. 

Incorporating retrospectives after key academic milestones could encourage reflection and continuous improvement. Maintaining adaptability in task management is also essential. Teams can reprioritize their backlog as workloads shift and postpone less critical tasks until lighter periods of the semester. Rotating the Scrum Master role can distribute leadership more evenly by allowing everyone to develop organizational and communication skills. During sprint planning, team members need to share any upcoming academic commitments to ensure that work is distributed fairly. Incorporating school terms and breaks as milestones, along with academic deliverables in sprint goals, helps integrate coursework and project progress. Over time, keeping a record of what adjustments worked well can serve as a valuable resource for future teams.
\subsection{Validity}\label{TV}

Our work is subject to several threats to validity. In this section, we reflect on the limitations related to internal, construct, and external validity and outline the measures we took to mitigate them.

\subsubsection{Internal Validity}

A first threat to internal validity arises from the authors' dual role as both researchers and active members of the development team. This insider position provided rich access to day-to-day interactions. Still, it may also have introduced bias in data interpretation and selective attention to events that fit pre-existing beliefs about the team. In addition, the presence of an ongoing study may have influenced behaviour (a Hawthorne effect \cite{sedgwick2015understanding}), with team members potentially communicating or contributing differently because they knew that their activity might later be analyzed.

Several data sources also involved manual processing. Discord messages and Notion tasks were inspected and coded by the authors, for example, when distinguishing project-related messages, annotating questions and answers, or linking tasks to commits. These activities are susceptible to misclassification and inconsistent judgment, particularly in cases that are borderline. We attempted to mitigate this by cross-checking interpretations and relying on clearly defined categories, but we did not employ formal inter-rater reliability measures. 
Finally, the study did not include a control group or comparison with other teams, which limits our ability to attribute observed patterns (e.g., stable productivity during remote periods) to specific practices or tools rather than to other unobserved factors such as individual diligence or prior team cohesion.

\subsubsection{Construct Validity}

Our Communication Effectiveness Index operationalizes communication quality through four dimensions: Activity, Responsiveness, Clarity, and Outcome. These dimensions were chosen because they mirror aspects frequently highlighted in prior work on software coordination and teamwork, including participation, timely replies, understandable and relevant exchanges, and the successful execution of tasks. However, they remain a simplification of the broader concept of "effective communication."

Some limitations follow from this operationalization. First, sentiment, relevance, and clarity judgments were performed manually, which introduces subjectivity and potential bias. Second, the thresholds for mapping raw metrics (such as response time or the proportion of unanswered questions) to normalized scores were defined by the team for this specific context, rather than being derived from large-scale empirical benchmarks. Different thresholds might yield different index values without necessarily changing the underlying communication behaviour. Alternative or complementary metrics, such as the network structure of interactions, richer sentiment analysis, or survey-based perceptions of communication quality, could provide a more nuanced picture.

\subsubsection{External Validity}

Our study focuses on a single student team in one university, working on a specific type of project within a single academic program. The team members shared a similar educational background, were of a comparable age, and often knew each other beforehand. These characteristics likely contributed to a cohesive working environment and may not be as effective in more heterogeneous or less familiar groups.

Moreover, the context is educational rather than industrial. While the team adopted Scrum-like practices and used widely available tools, their constraints, motivations, and stakes differ from those of professional software developers. For example, examination periods and semester boundaries significantly influenced the sprint cadence and workload. In contrast, commercial teams might be more strongly driven by customer deadlines or market pressures. Similarly, our hybrid setup involved a small group alternating between on-site and remote work, which differs from large distributed organizations with multiple time zones and departments. As a result, we expect some of the observed patterns, such as the usefulness of pause weeks around exams or the specific balance of meetings and asynchronous communication, to generalize primarily to similar student project settings rather than to all agile teams.

\subsection{Implications for Practice and Research}\label{IM}

This section summarizes the practical lessons that can be drawn from our case study and outlines how they can inform teaching practice, student teamwork, and future research.

\subsubsection{Implications for Instructors and Course Designers}

Our case study suggests that Scrum can be successfully integrated into long-running, project-based courses when its structure can be aligned with the students' academic calendar. Instead of prescribing fixed, industry-style sprint lengths, instructors can support student teams in selecting and justifying a cadence that fits their workload, such as one-week iterations with planned pause weeks around examinations. Making exam periods and institutional deadlines part of the official project timeline, for example, by treating them as milestones with reduced development expectations, helps normalize fluctuating velocity rather than framing it as a failure of discipline.

Our findings also highlight the importance of providing a minimal but coherent collaboration stack. Notion, Discord, and GitHub together covered task management, communication, and version control in a way that matched the team’s needs and introvert-leaning working style. Instructors designing similar courses can recommend a small set of integrated tools and provide templates, such as Notion boards for backlogs and meeting notes, or Discord channel structures, that lower the barrier to effective coordination without overconstraining teams. Finally, assessment schemes can take into account both artefact quality and process evidence (e.g., commit history, task boards, and communication patterns), encouraging students to treat the process as a first-class learning outcome rather than a formality.

\subsubsection{Implications for Student Teams}

For student teams, the project emphasizes the importance of explicitly negotiating how Scrum will be adapted to their specific constraints and preferences, rather than adopting it "by the book." Agreeing on realistic sprint goals, openly sharing upcoming academic commitments during planning, and using pause weeks strategically can reduce last-minute stress and prevent feelings of failure when progress slows.

The findings also show that role clarity and self-awareness about working styles matter as much as tool choice. Clearly defined responsibilities for backlog ownership, facilitation, and technical leadership supported psychological safety and kept the team moving forward. Teams may benefit from early conversations about personalities and preferences, such as whether members prefer brief, task-oriented meetings or more exploratory discussions. Additionally, they can benefit from selecting ceremonies and tools that align with those preferences. 

\subsubsection{Implications for Researchers}

From a research perspective, the study contributes an example of how communication and coordination in student Scrum teams can be operationalized using readily available artefacts. The Communication Effectiveness Index, based on activity, responsiveness, clarity, and outcomes, provides a simple and interpretable metric that other researchers can adapt to different tools or contexts. Future work could refine this index, validate it across multiple teams, or correlate it with additional outcomes such as perceived learning, code quality, or retention.

The project also highlights the need for more longitudinal studies of agile practices in education that extend beyond single-semester courses and capture how teams evolve across multiple phases, milestones, and academic cycles. Comparative research across teams with different tool stacks, sprint structures, or personality compositions could help disentangle which adaptations are most beneficial under which conditions. 

\subsection{Future Work}
In future studies, we plan to incorporate additional numerical methods to assess the effectiveness of collaboration tools, rather than relying solely on manual checks of messages and activities. Automated metrics, such as the speed at which people respond to messages, the frequency of task completion, the connectivity between different platforms (for example, GitHub commits linked to Notion tasks), and the density of the interaction network, can provide more transparent and more scalable insights into team communication. Using these data-based measures would also help us track collaboration over time and make the evaluation process less subjective.

\section{Conclusion}\label{C}

The development of a virtual reality project highlights how agile methodologies can be effectively applied in academic environments. The project demonstrated that when properly structured, Scrum principles enhance collaboration, accountability, and iterative learning even within the constraints of semester schedules and hybrid settings. Despite time-zone and workload challenges, the team successfully maintained stable communication and productivity by utilizing integrated digital tools, including Notion, Discord, and GitHub. These tools not only supported effective task management but also encouraged transparency, reflection, and autonomous participation, which are central to both agile and pedagogical goals.

Another key outcome of this study was the observation that hybrid or remote work did not necessarily hinder collaboration. Instead, it often enhances structured communication and asynchronous flexibility, indicating that digitally mediated teamwork can sustain engagement when social and technical systems are well-aligned. Moreover, the project reaffirmed the importance of human factors, particularly motivation, shared knowledge, and compatible personality traits, in determining team success.

In conclusion, this project demonstrates that combining immersive technology, agile methodologies, and teamwork facilitates students' learning and growth. When education mixes real-world software practices with creative work, students are better prepared for modern workplaces. Future studies should measure how these approaches affect learning, teamwork, and teaching over time.
Patterns observed here hold across different contexts, courses, and project types, providing stronger evidence on which Scrum adaptations are most robust.

\section*{Aknowledgements}
This research was financially supported by the TUM Campus Heilbronn \textit{Incentive Fund 2024} of the Technical University of Munich, TUM Campus Heilbronn. We gratefully acknowledge their support, which provided the essential resources and opportunities to conduct this study. 

\printcredits

%\bibliographystyle{cas-model2-names}

%\bibliography{00Paper}

\begin{thebibliography}{65}
\expandafter\ifx\csname natexlab\endcsname\relax\def\natexlab#1{#1}\fi
\providecommand{\url}[1]{\texttt{#1}}
\providecommand{\href}[2]{#2}
\providecommand{\path}[1]{#1}
\providecommand{\DOIprefix}{doi:}
\providecommand{\ArXivprefix}{arXiv:}
\providecommand{\URLprefix}{URL: }
\providecommand{\Pubmedprefix}{pmid:}
\providecommand{\doi}[1]{\href{http://dx.doi.org/#1}{\path{#1}}}
\providecommand{\Pubmed}[1]{\href{pmid:#1}{\path{#1}}}
\providecommand{\bibinfo}[2]{#2}
\ifx\xfnm\relax \def\xfnm[#1]{\unskip,\space#1}\fi
%Type = Article
\bibitem[{Alami and Krancher(2022)}]{alami2022scrum}
\bibinfo{author}{Alami, A.}, \bibinfo{author}{Krancher, O.},
  \bibinfo{year}{2022}.
\newblock \bibinfo{title}{How scrum adds value to achieving software quality?}
\newblock \bibinfo{journal}{Empirical Software Engineering}
  \bibinfo{volume}{27}, \bibinfo{pages}{165}.
%Type = Article
\bibitem[{AlGhamdi(2025)}]{alghamdi2025bridging}
\bibinfo{author}{AlGhamdi, R.}, \bibinfo{year}{2025}.
\newblock \bibinfo{title}{Bridging learning gaps through discord: peer-to-peer
  learning in computer graphics education}.
\newblock \bibinfo{journal}{Learning and Teaching in Higher Education: Gulf
  Perspectives} .
%Type = Article
\bibitem[{Amin et~al.(2020)Amin, Basri, Rehman, Capretz, Akbar, Gilal and
  Shabbir}]{amin2020impact}
\bibinfo{author}{Amin, A.}, \bibinfo{author}{Basri, S.},
  \bibinfo{author}{Rehman, M.}, \bibinfo{author}{Capretz, L.F.},
  \bibinfo{author}{Akbar, R.}, \bibinfo{author}{Gilal, A.R.},
  \bibinfo{author}{Shabbir, M.F.}, \bibinfo{year}{2020}.
\newblock \bibinfo{title}{The impact of personality traits and knowledge
  collection behavior on programmer creativity}.
\newblock \bibinfo{journal}{Information and Software Technology}
  \bibinfo{volume}{128}, \bibinfo{pages}{106405}.
%Type = Article
\bibitem[{de~Andrade et~al.(2024)de~Andrade, Jackson, Prikladnicki and van~der
  Hoek}]{de2024meetings}
\bibinfo{author}{de~Andrade, A.S.}, \bibinfo{author}{Jackson, V.},
  \bibinfo{author}{Prikladnicki, R.}, \bibinfo{author}{van~der Hoek, A.},
  \bibinfo{year}{2024}.
\newblock \bibinfo{title}{On meetings involving remote software teams: A
  systematic literature review}.
\newblock \bibinfo{journal}{Information and Software Technology}
  \bibinfo{volume}{175}, \bibinfo{pages}{107541}.
%Type = Misc
\bibitem[{Andree et~al.(2025)Andree, Berrezueta-Guzman, Krusche, Pufahl and
  Wagner}]{andree2025}
\bibinfo{author}{Andree, K.}, \bibinfo{author}{Berrezueta-Guzman, S.},
  \bibinfo{author}{Krusche, S.}, \bibinfo{author}{Pufahl, L.},
  \bibinfo{author}{Wagner, S.}, \bibinfo{year}{2025}.
\newblock \bibinfo{title}{How soft skills shape first-year success in higher
  education}.
\newblock \URLprefix \url{https://arxiv.org/abs/2505.21696},
  \href{http://arxiv.org/abs/2505.21696}{\tt arXiv:2505.21696}.
%Type = Article
\bibitem[{Aragon{\'e}s-Jeric{\'o} and Canales-Ronda(2022)}]{aragones2022agile}
\bibinfo{author}{Aragon{\'e}s-Jeric{\'o}, C.}, \bibinfo{author}{Canales-Ronda,
  P.}, \bibinfo{year}{2022}.
\newblock \bibinfo{title}{Agile learning in marketing: Scrum in higher
  education}.
\newblock \bibinfo{journal}{Journal of Management and Business Education}
  \bibinfo{volume}{5}, \bibinfo{pages}{345--360}.
%Type = Article
\bibitem[{Ask~Uggla(2024)}]{ask2024hybrid}
\bibinfo{author}{Ask~Uggla, M.}, \bibinfo{year}{2024}.
\newblock \bibinfo{title}{Hybrid work within agile software development teams}.
\newblock \bibinfo{journal}{LU-CS-EX} .
%Type = Inproceedings
\bibitem[{Barroso et~al.(2017)Barroso, Madureira, Soares and
  do~Nascimento}]{barroso2017influence}
\bibinfo{author}{Barroso, A.S.}, \bibinfo{author}{Madureira, J.S.},
  \bibinfo{author}{Soares, M.S.}, \bibinfo{author}{do~Nascimento, R.P.},
  \bibinfo{year}{2017}.
\newblock \bibinfo{title}{Influence of human personality in software
  engineering-a systematic literature review}, in:
  \bibinfo{booktitle}{International Conference on Enterprise Information
  Systems}, \bibinfo{organization}{SciTePress}. pp. \bibinfo{pages}{53--62}.
%Type = Article
\bibitem[{Beecham et~al.(2008)Beecham, Baddoo, Hall, Robinson and
  Sharp}]{beecham2008motivation}
\bibinfo{author}{Beecham, S.}, \bibinfo{author}{Baddoo, N.},
  \bibinfo{author}{Hall, T.}, \bibinfo{author}{Robinson, H.},
  \bibinfo{author}{Sharp, H.}, \bibinfo{year}{2008}.
\newblock \bibinfo{title}{Motivation in software engineering: A systematic
  literature review}.
\newblock \bibinfo{journal}{Information and software technology}
  \bibinfo{volume}{50}, \bibinfo{pages}{860--878}.
%Type = Inproceedings
\bibitem[{Berrezueta-Guzman et~al.(2024)Berrezueta-Guzman, Parmacli, Krusche
  and Wagner}]{berrezueta2024interactive}
\bibinfo{author}{Berrezueta-Guzman, S.}, \bibinfo{author}{Parmacli, I.},
  \bibinfo{author}{Krusche, S.}, \bibinfo{author}{Wagner, S.},
  \bibinfo{year}{2024}.
\newblock \bibinfo{title}{Interactive learning in computer science education
  supported by a discord chatbot}, in: \bibinfo{booktitle}{2024 IEEE 3rd German
  Education Conference (GECon)}, \bibinfo{organization}{IEEE}. pp.
  \bibinfo{pages}{1--6}.
%Type = Techreport
\bibitem[{Bloom et~al.(2022)Bloom, Han and Liang}]{bloom2022hybrid}
\bibinfo{author}{Bloom, N.}, \bibinfo{author}{Han, R.}, \bibinfo{author}{Liang,
  J.}, \bibinfo{year}{2022}.
\newblock \bibinfo{title}{How hybrid working from home works out}.
\newblock \bibinfo{type}{Technical Report}. National Bureau of economic
  research.
%Type = Inproceedings
\bibitem[{Bridson et~al.(2022)Bridson, Atkinson and
  Fleming}]{bridson2022delivering}
\bibinfo{author}{Bridson, K.}, \bibinfo{author}{Atkinson, J.},
  \bibinfo{author}{Fleming, S.D.}, \bibinfo{year}{2022}.
\newblock \bibinfo{title}{Delivering round-the-clock help to software
  engineering students using discord: An experience report}, in:
  \bibinfo{booktitle}{Proceedings of the 53rd ACM Technical Symposium on
  Computer Science Education-Volume 1}, pp. \bibinfo{pages}{759--765}.
%Type = Article
\bibitem[{Capretz(2003)}]{capretz2003personality}
\bibinfo{author}{Capretz, L.F.}, \bibinfo{year}{2003}.
\newblock \bibinfo{title}{Personality types in software engineering}.
\newblock \bibinfo{journal}{International Journal of Human-Computer Studies}
  \bibinfo{volume}{58}, \bibinfo{pages}{207--214}.
%Type = Article
\bibitem[{Capretz et~al.(2015)Capretz, Varona and Raza}]{capretz2015influence}
\bibinfo{author}{Capretz, L.F.}, \bibinfo{author}{Varona, D.},
  \bibinfo{author}{Raza, A.}, \bibinfo{year}{2015}.
\newblock \bibinfo{title}{Influence of personality types in software tasks
  choices}.
\newblock \bibinfo{journal}{Computers in Human behavior} \bibinfo{volume}{52},
  \bibinfo{pages}{373--378}.
%Type = Inproceedings
\bibitem[{Carvalho et~al.(2018)Carvalho, Fernandes, Lima and
  Mesquita}]{carvalho2018making}
\bibinfo{author}{Carvalho, J.D.}, \bibinfo{author}{Fernandes, S.},
  \bibinfo{author}{Lima, R.M.}, \bibinfo{author}{Mesquita, D.},
  \bibinfo{year}{2018}.
\newblock \bibinfo{title}{Making pbl teams more effective with scrum}, in:
  \bibinfo{booktitle}{International Symposium on Project Approaches in
  Engineering Education}, pp. \bibinfo{pages}{64--72}.
%Type = Article
\bibitem[{Chandrasekar(2011)}]{chandrasekar2011impact}
\bibinfo{author}{Chandrasekar, L.}, \bibinfo{year}{2011}.
\newblock \bibinfo{title}{The impact of collaboration tools on student
  engagement}.
\newblock \bibinfo{journal}{Morning Watch} \bibinfo{volume}{5},
  \bibinfo{pages}{1--18}.
%Type = Article
\bibitem[{Chowdhury et~al.(2025)Chowdhury, Sultana, Chowdhury and
  Chowdhury}]{chowdhury2025agile}
\bibinfo{author}{Chowdhury, B.}, \bibinfo{author}{Sultana, N.},
  \bibinfo{author}{Chowdhury, S.}, \bibinfo{author}{Chowdhury, S.},
  \bibinfo{year}{2025}.
\newblock \bibinfo{title}{Agile approach to enhance student's capstone
  (industry-based) product delivery in tertiary education}.
\newblock \bibinfo{journal}{WSEAS Transactions on Advances in Engineering
  Education} \bibinfo{volume}{22}, \bibinfo{pages}{17--32}.
%Type = Inproceedings
\bibitem[{Christensen et~al.(2025)Christensen, Paasivaara and
  Salman}]{christensen2025hybrid}
\bibinfo{author}{Christensen, E.L.}, \bibinfo{author}{Paasivaara, M.},
  \bibinfo{author}{Salman, I.}, \bibinfo{year}{2025}.
\newblock \bibinfo{title}{Hybrid work in agile software development: Recurring
  meetings}, in: \bibinfo{booktitle}{2025 IEEE/ACM 18th International
  Conference on Cooperative and Human Aspects of Software Engineering (CHASE)},
  \bibinfo{organization}{IEEE}. pp. \bibinfo{pages}{120--130}.
%Type = Article
\bibitem[{{\v{C}}isar et~al.(2025){\v{C}}isar, Pinter, {\v{C}}isar and
  {\DJ}ikanovi{\'c}}]{vcisar2025impact}
\bibinfo{author}{{\v{C}}isar, S.M.}, \bibinfo{author}{Pinter, R.},
  \bibinfo{author}{{\v{C}}isar, P.}, \bibinfo{author}{{\DJ}ikanovi{\'c}, P.},
  \bibinfo{year}{2025}.
\newblock \bibinfo{title}{The impact of scrum methodology on student motivation
  and problem-solving skills.}
\newblock \bibinfo{journal}{Interdisciplinary Description of Complex Systems}
  \bibinfo{volume}{23}.
%Type = Article
\bibitem[{Cruz et~al.(2015)Cruz, Da~Silva and Capretz}]{cruz2015forty}
\bibinfo{author}{Cruz, S.}, \bibinfo{author}{Da~Silva, F.Q.},
  \bibinfo{author}{Capretz, L.F.}, \bibinfo{year}{2015}.
\newblock \bibinfo{title}{Forty years of research on personality in software
  engineering: A mapping study}.
\newblock \bibinfo{journal}{Computers in Human Behavior} \bibinfo{volume}{46},
  \bibinfo{pages}{94--113}.
%Type = Inproceedings
\bibitem[{Culas(2025)}]{culas2025advancing}
\bibinfo{author}{Culas, F.}, \bibinfo{year}{2025}.
\newblock \bibinfo{title}{Advancing cognitive inclusivity in software
  engineering tools and practices}, in: \bibinfo{booktitle}{2025 IEEE/ACM 47th
  International Conference on Software Engineering: Companion Proceedings
  (ICSE-Companion)}, \bibinfo{organization}{IEEE}. pp.
  \bibinfo{pages}{166--168}.
%Type = Inproceedings
\bibitem[{Dabbish et~al.(2012)Dabbish, Stuart, Tsay and
  Herbsleb}]{dabbish2012social}
\bibinfo{author}{Dabbish, L.}, \bibinfo{author}{Stuart, C.},
  \bibinfo{author}{Tsay, J.}, \bibinfo{author}{Herbsleb, J.},
  \bibinfo{year}{2012}.
\newblock \bibinfo{title}{Social coding in github: transparency and
  collaboration in an open software repository}, in:
  \bibinfo{booktitle}{Proceedings of the ACM 2012 conference on computer
  supported cooperative work}, pp. \bibinfo{pages}{1277--1286}.
%Type = Article
\bibitem[{Desch{\^e}nes(2024)}]{deschenes2024digital}
\bibinfo{author}{Desch{\^e}nes, A.A.}, \bibinfo{year}{2024}.
\newblock \bibinfo{title}{Digital literacy, the use of collaborative
  technologies, and perceived social proximity in a hybrid work environment:
  Technology as a social binder}.
\newblock \bibinfo{journal}{Computers in Human Behavior Reports}
  \bibinfo{volume}{13}, \bibinfo{pages}{100351}.
%Type = Article
\bibitem[{Endriulaitien{\.e} and
  Cirtautien{\.e}(2021)}]{endriulaitiene2021team}
\bibinfo{author}{Endriulaitien{\.e}, A.}, \bibinfo{author}{Cirtautien{\.e},
  L.}, \bibinfo{year}{2021}.
\newblock \bibinfo{title}{Team effectiveness in software development: the role
  of personality and work factors}.
\newblock \bibinfo{journal}{Business: Theory and Practice}
  \bibinfo{volume}{22}, \bibinfo{pages}{55--68}.
%Type = Article
\bibitem[{Fernandes et~al.(2021)Fernandes, Dinis-Carvalho and
  Ferreira-Oliveira}]{fernandes2021improving}
\bibinfo{author}{Fernandes, S.}, \bibinfo{author}{Dinis-Carvalho, J.},
  \bibinfo{author}{Ferreira-Oliveira, A.T.}, \bibinfo{year}{2021}.
\newblock \bibinfo{title}{Improving the performance of student teams in
  project-based learning with scrum}.
\newblock \bibinfo{journal}{Education sciences} \bibinfo{volume}{11},
  \bibinfo{pages}{444}.
%Type = Misc
\bibitem[{{GitHub, Inc.}(2025)}]{github2025}
\bibinfo{author}{{GitHub, Inc.}}, \bibinfo{year}{2025}.
\newblock \bibinfo{title}{About github}.
\newblock \URLprefix \url{https://github.com/about}. \bibinfo{note}{accessed
  October 24, 2025}.
%Type = Article
\bibitem[{Goldberg()}]{goldbergstory}
\bibinfo{author}{Goldberg, A.}, .
\newblock \bibinfo{title}{Story telling: Collaborative software engineering}.
\newblock \bibinfo{journal}{Journal of Object Technology, http://www. jot.
  fm/issues/issue\_2002\_05/column1} .
%Type = Article
\bibitem[{Graziotin et~al.(2018)Graziotin, Fagerholm, Wang and
  Abrahamsson}]{graziotin2018happens}
\bibinfo{author}{Graziotin, D.}, \bibinfo{author}{Fagerholm, F.},
  \bibinfo{author}{Wang, X.}, \bibinfo{author}{Abrahamsson, P.},
  \bibinfo{year}{2018}.
\newblock \bibinfo{title}{What happens when software developers are (un)
  happy}.
\newblock \bibinfo{journal}{Journal of Systems and Software}
  \bibinfo{volume}{140}, \bibinfo{pages}{32--47}.
%Type = Inproceedings
\bibitem[{Gustavsson(2016)}]{gustavsson2016benefits}
\bibinfo{author}{Gustavsson, T.}, \bibinfo{year}{2016}.
\newblock \bibinfo{title}{Benefits of agile project management in a
  non-software development context: A literature review}, in:
  \bibinfo{booktitle}{Fifth International Scientific Conference on Project
  Management in the Baltic Countries, April 14-15, 2016, Riga, University of
  Latvia}, \bibinfo{organization}{Latvijas Universitate}. pp.
  \bibinfo{pages}{114--124}.
%Type = Article
\bibitem[{Handke et~al.(2024)Handke, Aldana, Costa and
  O’Neill}]{handke2024hybrid}
\bibinfo{author}{Handke, L.}, \bibinfo{author}{Aldana, A.},
  \bibinfo{author}{Costa, P.L.}, \bibinfo{author}{O’Neill, T.A.},
  \bibinfo{year}{2024}.
\newblock \bibinfo{title}{Hybrid teamwork: What we know and where we can go
  from here}.
\newblock \bibinfo{journal}{Small Group Research} \bibinfo{volume}{55},
  \bibinfo{pages}{805--835}.
%Type = Article
\bibitem[{Herbsleb and Mockus(2003)}]{herbsleb2003empirical}
\bibinfo{author}{Herbsleb, J.D.}, \bibinfo{author}{Mockus, A.},
  \bibinfo{year}{2003}.
\newblock \bibinfo{title}{An empirical study of speed and communication in
  globally distributed software development}.
\newblock \bibinfo{journal}{IEEE Transactions on software engineering}
  \bibinfo{volume}{29}, \bibinfo{pages}{481--494}.
%Type = Article
\bibitem[{Hidalgo(2019)}]{hidalgo2019adapting}
\bibinfo{author}{Hidalgo, E.S.}, \bibinfo{year}{2019}.
\newblock \bibinfo{title}{Adapting the scrum framework for agile project
  management in science: case study of a distributed research initiative}.
\newblock \bibinfo{journal}{Heliyon} \bibinfo{volume}{5}.
%Type = Article
\bibitem[{Hidellaarachchi et~al.(2024)Hidellaarachchi, Grundy, Hoda and
  Mueller}]{hidellaarachchi2024impact}
\bibinfo{author}{Hidellaarachchi, D.}, \bibinfo{author}{Grundy, J.},
  \bibinfo{author}{Hoda, R.}, \bibinfo{author}{Mueller, I.},
  \bibinfo{year}{2024}.
\newblock \bibinfo{title}{The impact of personality on requirements engineering
  activities: A mixed-methods study}.
\newblock \bibinfo{journal}{Empirical Software Engineering}
  \bibinfo{volume}{29}, \bibinfo{pages}{32}.
%Type = Article
\bibitem[{Hussein and Hassan(2025)}]{hussein2025collaboration}
\bibinfo{author}{Hussein, R.M.}, \bibinfo{author}{Hassan, B.A.},
  \bibinfo{year}{2025}.
\newblock \bibinfo{title}{Collaboration tools and their role in agile software
  projects}.
\newblock \bibinfo{journal}{arXiv preprint arXiv:2506.10985} .
%Type = Article
\bibitem[{Jackson et~al.(2022)Jackson, van~der Hoek, Prikladnicki and
  Ebert}]{jackson2022collaboration}
\bibinfo{author}{Jackson, V.}, \bibinfo{author}{van~der Hoek, A.},
  \bibinfo{author}{Prikladnicki, R.}, \bibinfo{author}{Ebert, C.},
  \bibinfo{year}{2022}.
\newblock \bibinfo{title}{Collaboration tools for developers}.
\newblock \bibinfo{journal}{IEEE software} \bibinfo{volume}{39},
  \bibinfo{pages}{7--15}.
%Type = Article
\bibitem[{Kadenic et~al.(2023)Kadenic, Koumaditis and
  Junker-Jensen}]{kadenic2023mastering}
\bibinfo{author}{Kadenic, M.D.}, \bibinfo{author}{Koumaditis, K.},
  \bibinfo{author}{Junker-Jensen, L.}, \bibinfo{year}{2023}.
\newblock \bibinfo{title}{Mastering scrum with a focus on team maturity and key
  components of scrum}.
\newblock \bibinfo{journal}{Information and Software Technology}
  \bibinfo{volume}{153}, \bibinfo{pages}{107079}.
%Type = Inproceedings
\bibitem[{Khanna et~al.(2024a)Khanna, Christensen, Gosu, Wang and
  Paasivaara}]{khanna2024hybrid}
\bibinfo{author}{Khanna, D.}, \bibinfo{author}{Christensen, E.L.},
  \bibinfo{author}{Gosu, S.}, \bibinfo{author}{Wang, X.},
  \bibinfo{author}{Paasivaara, M.}, \bibinfo{year}{2024}a.
\newblock \bibinfo{title}{Hybrid work meets agile software development: A
  systematic mapping study}, in: \bibinfo{booktitle}{Proceedings of the 2024
  IEEE/ACM 17th International Conference on Cooperative and Human Aspects of
  Software Engineering}, pp. \bibinfo{pages}{57--67}.
%Type = Inproceedings
\bibitem[{Khanna et~al.(2024b)Khanna, Edison, Nguyen-Duc and
  Kemell}]{khanna2024software}
\bibinfo{author}{Khanna, D.}, \bibinfo{author}{Edison, H.},
  \bibinfo{author}{Nguyen-Duc, A.}, \bibinfo{author}{Kemell, K.K.},
  \bibinfo{year}{2024}b.
\newblock \bibinfo{title}{Software companies' responses to hybrid working}, in:
  \bibinfo{booktitle}{2024 50th Euromicro Conference on Software Engineering
  and Advanced Applications (SEAA)}, \bibinfo{organization}{IEEE}. pp.
  \bibinfo{pages}{244--251}.
%Type = Article
\bibitem[{Kim et~al.(2025)Kim, Klein-Balajee, Kelly and
  Hiniker}]{kim2025discord}
\bibinfo{author}{Kim, J.}, \bibinfo{author}{Klein-Balajee, T.},
  \bibinfo{author}{Kelly, R.M.}, \bibinfo{author}{Hiniker, A.},
  \bibinfo{year}{2025}.
\newblock \bibinfo{title}{Discord's design encourages" third place" social
  media experiences}.
\newblock \bibinfo{journal}{arXiv preprint arXiv:2501.09951} .
%Type = Article
\bibitem[{Lenberga et~al.()Lenberga, Feldtb and Wallgrenc}]{lenbergabehavioral}
\bibinfo{author}{Lenberga, P.}, \bibinfo{author}{Feldtb, R.},
  \bibinfo{author}{Wallgrenc, L.G.}, .
\newblock \bibinfo{title}{Behavioral software engineering: a definition and
  systematic} .
%Type = Article
\bibitem[{Masood et~al.(2018)Masood, Hoda and Blincoe}]{masood2018adapting}
\bibinfo{author}{Masood, Z.}, \bibinfo{author}{Hoda, R.},
  \bibinfo{author}{Blincoe, K.}, \bibinfo{year}{2018}.
\newblock \bibinfo{title}{Adapting agile practices in university contexts}.
\newblock \bibinfo{journal}{Journal of Systems and Software}
  \bibinfo{volume}{144}, \bibinfo{pages}{501--510}.
%Type = Article
\bibitem[{Meyer et~al.(2019)Meyer, Barr, Bird and Zimmermann}]{meyer2019today}
\bibinfo{author}{Meyer, A.N.}, \bibinfo{author}{Barr, E.T.},
  \bibinfo{author}{Bird, C.}, \bibinfo{author}{Zimmermann, T.},
  \bibinfo{year}{2019}.
\newblock \bibinfo{title}{Today was a good day: The daily life of software
  developers}.
\newblock \bibinfo{journal}{IEEE Transactions on Software Engineering}
  \bibinfo{volume}{47}, \bibinfo{pages}{863--880}.
%Type = Article
\bibitem[{Miller et~al.()Miller, Mellinger and Yasar}]{millerhuman}
\bibinfo{author}{Miller, S.}, \bibinfo{author}{Mellinger, A.},
  \bibinfo{author}{Yasar, H.}, .
\newblock \bibinfo{title}{Human factors in software engineering} .
%Type = Article
\bibitem[{Nguyen-Duc et~al.(2024)Nguyen-Duc, Khanna, Le, Greer, Wang, Zaina,
  Matturro, Melegati, Guerra, Kettunen et~al.}]{nguyen2024work}
\bibinfo{author}{Nguyen-Duc, A.}, \bibinfo{author}{Khanna, D.},
  \bibinfo{author}{Le, G.H.}, \bibinfo{author}{Greer, D.},
  \bibinfo{author}{Wang, X.}, \bibinfo{author}{Zaina, L.M.},
  \bibinfo{author}{Matturro, G.}, \bibinfo{author}{Melegati, J.},
  \bibinfo{author}{Guerra, E.}, \bibinfo{author}{Kettunen, P.}, et~al.,
  \bibinfo{year}{2024}.
\newblock \bibinfo{title}{Work-from-home impacts on software project: A global
  study on software development practices and stakeholder perceptions}.
\newblock \bibinfo{journal}{Software: Practice and Experience}
  \bibinfo{volume}{54}, \bibinfo{pages}{896--926}.
%Type = Inproceedings
\bibitem[{Persson et~al.(2011)Persson, Kruzela, Allder, Johansson and
  Johansson}]{persson2011use}
\bibinfo{author}{Persson, M.}, \bibinfo{author}{Kruzela, I.},
  \bibinfo{author}{Allder, K.}, \bibinfo{author}{Johansson, O.},
  \bibinfo{author}{Johansson, P.}, \bibinfo{year}{2011}.
\newblock \bibinfo{title}{On the use of scrum in project driven higher
  education}, in: \bibinfo{booktitle}{Proceedings of the International
  Conference on Frontiers in Education: Computer Science and Computer
  Engineering (FECS)}, \bibinfo{organization}{The Steering Committee of The
  World Congress in Computer Science, Computer~…}. p.~\bibinfo{pages}{1}.
%Type = Inproceedings
\bibitem[{Pozdniakov et~al.(2025)Pozdniakov, Brazil, Poquet, Krusche,
  Berrezueta-Guzman, Sadiq and Khosravi}]{pozdniakov2025misunderstandings}
\bibinfo{author}{Pozdniakov, S.}, \bibinfo{author}{Brazil, J.},
  \bibinfo{author}{Poquet, O.}, \bibinfo{author}{Krusche, S.},
  \bibinfo{author}{Berrezueta-Guzman, S.}, \bibinfo{author}{Sadiq, S.},
  \bibinfo{author}{Khosravi, H.}, \bibinfo{year}{2025}.
\newblock \bibinfo{title}{From misunderstandings to learning opportunities:
  Leveraging generative ai in discussion forums to support student learning},
  in: \bibinfo{booktitle}{International Conference on Artificial Intelligence
  in Education}, \bibinfo{organization}{Springer}. pp.
  \bibinfo{pages}{291--298}.
%Type = Article
\bibitem[{Priyanka(2025)}]{priyanka2025analyzing}
\bibinfo{author}{Priyanka, M.}, \bibinfo{year}{2025}.
\newblock \bibinfo{title}{Analyzing the impact of agile methodologies on
  software quality and delivery speed: A comparative study}.
\newblock \bibinfo{journal}{World Journal of Advanced Research and Reviews}
  \bibinfo{volume}{25}, \bibinfo{pages}{1207--1216}.
%Type = Inproceedings
\bibitem[{Raglianti et~al.(2022)Raglianti, Nagy, Minelli and
  Lanza}]{raglianti2022using}
\bibinfo{author}{Raglianti, M.}, \bibinfo{author}{Nagy, C.},
  \bibinfo{author}{Minelli, R.}, \bibinfo{author}{Lanza, M.},
  \bibinfo{year}{2022}.
\newblock \bibinfo{title}{Using discord conversations as program comprehension
  aid}, in: \bibinfo{booktitle}{Proceedings of the 30th IEEE/ACM International
  Conference on Program Comprehension}, pp. \bibinfo{pages}{597--601}.
%Type = Article
\bibitem[{Rajagopalan et~al.(2025)Rajagopalan, Woodside and
  Belanger}]{rajagopalan2025agile}
\bibinfo{author}{Rajagopalan, H.K.}, \bibinfo{author}{Woodside, S.},
  \bibinfo{author}{Belanger, K.L.}, \bibinfo{year}{2025}.
\newblock \bibinfo{title}{An agile approach to student consulting projects:
  Iteration and communication to improve decision making, presentations, and
  teamwork}.
\newblock \bibinfo{journal}{INFORMS Transactions on Education}
  \bibinfo{volume}{26}, \bibinfo{pages}{34--47}.
%Type = Article
\bibitem[{Rani and Suneja(2025)}]{rani2025attracting}
\bibinfo{author}{Rani, S.}, \bibinfo{author}{Suneja, A.}, \bibinfo{year}{2025}.
\newblock \bibinfo{title}{Attracting talent: understanding generation z’s
  expectations of technology-driven workplaces}.
\newblock \bibinfo{journal}{Vilakshan--XIMB Journal of Management} .
%Type = Article
\bibitem[{Redmond(2023)}]{redmond2023project}
\bibinfo{author}{Redmond, F.}, \bibinfo{year}{2023}.
\newblock \bibinfo{title}{Project management of an online change laboratory
  using notion}.
\newblock \bibinfo{journal}{Bureau de Change Laboratory} \bibinfo{volume}{1},
  \bibinfo{pages}{1--1}.
%Type = Inproceedings
\bibitem[{Reynolds et~al.(2023)Reynolds, Caldwell, Procko and
  Ochoa}]{reynolds2023scrum}
\bibinfo{author}{Reynolds, S.}, \bibinfo{author}{Caldwell, A.},
  \bibinfo{author}{Procko, T.}, \bibinfo{author}{Ochoa, O.},
  \bibinfo{year}{2023}.
\newblock \bibinfo{title}{Scrum in the classroom: An implementation guide}, in:
  \bibinfo{booktitle}{2023 IEEE Frontiers in Education Conference (FIE)},
  \bibinfo{organization}{IEEE}. pp. \bibinfo{pages}{01--08}.
%Type = Article
\bibitem[{Russo et~al.(2022)Russo, Masegosa and Stol}]{russo2022anecdote}
\bibinfo{author}{Russo, D.}, \bibinfo{author}{Masegosa, A.R.},
  \bibinfo{author}{Stol, K.J.}, \bibinfo{year}{2022}.
\newblock \bibinfo{title}{From anecdote to evidence: the relationship between
  personality and need for cognition of developers}.
\newblock \bibinfo{journal}{Empirical Software Engineering}
  \bibinfo{volume}{27}, \bibinfo{pages}{71}.
%Type = Article
\bibitem[{Saleh et~al.(2024)Saleh, Abbas, Latif and Khalil}]{saleh2024agile}
\bibinfo{author}{Saleh, H.H.}, \bibinfo{author}{Abbas, Z.A.},
  \bibinfo{author}{Latif, N.}, \bibinfo{author}{Khalil, Z.T.},
  \bibinfo{year}{2024}.
\newblock \bibinfo{title}{Agile management in healthcare improving patient
  outcomes through flexibility and responsiveness}.
\newblock \bibinfo{journal}{Journal of Ecohumanism} \bibinfo{volume}{3},
  \bibinfo{pages}{633--649}.
%Type = Article
\bibitem[{Santana et~al.(2025)Santana, Monte, de~Ara{\'u}jo~Silva, Carneiro,
  Freire, Santos and Mendon{\c{c}}a}]{santana2025psychological}
\bibinfo{author}{Santana, B.}, \bibinfo{author}{Monte, L.},
  \bibinfo{author}{de~Ara{\'u}jo~Silva, B.S.}, \bibinfo{author}{Carneiro, G.},
  \bibinfo{author}{Freire, S.}, \bibinfo{author}{Santos, J.A.M.},
  \bibinfo{author}{Mendon{\c{c}}a, M.}, \bibinfo{year}{2025}.
\newblock \bibinfo{title}{Psychological safety in software workplaces: A
  systematic literature review}.
\newblock \bibinfo{journal}{Information and Software Technology} ,
  \bibinfo{pages}{107838}.
%Type = Article
\bibitem[{Santana et~al.(2017)Santana, Santos, Silva, Villar, Rocha and
  Gon{\c{c}}alves}]{santana2017scrum}
\bibinfo{author}{Santana, L.F.}, \bibinfo{author}{Santos, L.F.C.d.},
  \bibinfo{author}{Silva, T.S.C.}, \bibinfo{author}{Villar, V.B.},
  \bibinfo{author}{Rocha, F.G.}, \bibinfo{author}{Gon{\c{c}}alves, V.},
  \bibinfo{year}{2017}.
\newblock \bibinfo{title}{Scrum as a platform to manage students in projects of
  technological development and scientific initiation: a study case realized at
  unit/se} .
%Type = Article
\bibitem[{Sarma(2005)}]{sarma2005survey}
\bibinfo{author}{Sarma, A.}, \bibinfo{year}{2005}.
\newblock \bibinfo{title}{A survey of collaborative tools in software
  development}.
\newblock \bibinfo{journal}{Institute for software research, Donald Bren school
  of information and computer science, University of California, Irvine} .
%Type = Article
\bibitem[{Sedgwick and Greenwood(2015)}]{sedgwick2015understanding}
\bibinfo{author}{Sedgwick, P.}, \bibinfo{author}{Greenwood, N.},
  \bibinfo{year}{2015}.
\newblock \bibinfo{title}{Understanding the hawthorne effect}.
\newblock \bibinfo{journal}{Bmj} \bibinfo{volume}{351}.
%Type = Article
\bibitem[{Soomro et~al.(2016)Soomro, Salleh, Mendes, Grundy, Burch and
  Nordin}]{soomro2016effect}
\bibinfo{author}{Soomro, A.B.}, \bibinfo{author}{Salleh, N.},
  \bibinfo{author}{Mendes, E.}, \bibinfo{author}{Grundy, J.},
  \bibinfo{author}{Burch, G.}, \bibinfo{author}{Nordin, A.},
  \bibinfo{year}{2016}.
\newblock \bibinfo{title}{The effect of software engineers’ personality
  traits on team climate and performance: a systematic literature} .
%Type = Inproceedings
\bibitem[{Subash et~al.(2022)Subash, Kumar, Vadlamani, Chatterjee and
  Baysal}]{subash2022disco}
\bibinfo{author}{Subash, K.M.}, \bibinfo{author}{Kumar, L.P.},
  \bibinfo{author}{Vadlamani, S.L.}, \bibinfo{author}{Chatterjee, P.},
  \bibinfo{author}{Baysal, O.}, \bibinfo{year}{2022}.
\newblock \bibinfo{title}{Disco: A dataset of discord chat conversations for
  software engineering research}, in: \bibinfo{booktitle}{Proceedings of the
  19th International Conference on Mining Software Repositories}, pp.
  \bibinfo{pages}{227--231}.
%Type = Article
\bibitem[{Umezurike et~al.(2025)Umezurike, Akinrinoye, Kufile, Abiodun
  Yusuf~Onifade, Otokiti and Ejike}]{umezurike2025review}
\bibinfo{author}{Umezurike, S.A.}, \bibinfo{author}{Akinrinoye, O.V.},
  \bibinfo{author}{Kufile, O.T.}, \bibinfo{author}{Abiodun Yusuf~Onifade, A.},
  \bibinfo{author}{Otokiti, B.O.}, \bibinfo{author}{Ejike, O.G.},
  \bibinfo{year}{2025}.
\newblock \bibinfo{title}{A review of agile marketing in cross-functional
  teams: Driving product growth through collaboration}.
\newblock \bibinfo{journal}{Journal of Frontiers in Multidisciplinary Research}
  \bibinfo{volume}{6}, \bibinfo{pages}{23--40}.
%Type = Inproceedings
\bibitem[{Vandersluis(2014)}]{vandersluis2014apply}
\bibinfo{author}{Vandersluis, C.}, \bibinfo{year}{2014}.
\newblock \bibinfo{title}{Apply agile methodology to non-software enterprise
  projects}, \bibinfo{organization}{Project Management Institute}.
%Type = Inproceedings
\bibitem[{Whitehead(2007)}]{whitehead2007collaboration}
\bibinfo{author}{Whitehead, J.}, \bibinfo{year}{2007}.
\newblock \bibinfo{title}{Collaboration in software engineering: A roadmap},
  in: \bibinfo{booktitle}{Future of Software Engineering (FOSE'07)},
  \bibinfo{organization}{IEEE}. pp. \bibinfo{pages}{214--225}.
%Type = Inproceedings
\bibitem[{Z{\"a}hl et~al.(2023)Z{\"a}hl, Theis, Wolf and
  K{\"o}hler}]{zahl2023teamwork}
\bibinfo{author}{Z{\"a}hl, P.M.}, \bibinfo{author}{Theis, S.},
  \bibinfo{author}{Wolf, M.R.}, \bibinfo{author}{K{\"o}hler, K.},
  \bibinfo{year}{2023}.
\newblock \bibinfo{title}{Teamwork in software development and what personality
  has to do with it-an overview}, in: \bibinfo{booktitle}{International
  Conference on Human-Computer Interaction}, \bibinfo{organization}{Springer}.
  pp. \bibinfo{pages}{130--153}.
%Type = Article
\bibitem[{Zolduoarrati et~al.(2023)Zolduoarrati, Licorish and
  Stanger}]{zolduoarrati2023secondary}
\bibinfo{author}{Zolduoarrati, E.}, \bibinfo{author}{Licorish, S.A.},
  \bibinfo{author}{Stanger, N.}, \bibinfo{year}{2023}.
\newblock \bibinfo{title}{Secondary studies on human aspects in software
  engineering: A tertiary study}.
\newblock \bibinfo{journal}{Journal of Systems and Software}
  \bibinfo{volume}{200}, \bibinfo{pages}{111654}.

\end{thebibliography}

\end{multicols}
\end{document}